\input{epsf}
\documentstyle[prd,eqsecnum,aps,twocolumn]{revtex}
\catcode`\@=11

\def\maketitle2{\par % Uses \twocolumn[\@maketitle2].
\begingroup
\let\cite\@bylinecite
\def\thefootnote{\fnsymbol{footnote}}%
\twocolumn[\@maketitle2\vskip2pc]%
\thispagestyle{plain}\@thanks
\endgroup
\def\thefootnote{\arabic{footnote}}%
\setcounter{footnote}{0}%
\let\maketitle2\relax \let\@maketitle2\relax
\let\@thanks\relax \let\@authoraddress\relax \let\@title\relax
\let\@date\relax \let\thanks\relax \let\@abstract\relax 
\let\@pacs\relax}

\def\abstract#1{\gdef\@abstract{{\par % Store abstract text. 
\bgroup
\ifdim\prevdepth=-1000pt \prevdepth0pt\fi
\hsize\columnwidth
\dimen0=-\prevdepth \advance\dimen0 by17.5pt \nointerlineskip
\small\vrule width 0pt height\dimen0 \relax}{~~}#1\egroup}}

\def\pacs#1{\gdef\@pacs{{\par % Store PACS numbers as \@pacs.
\bgroup
\hsize\columnwidth \parindent0pt
\ifdim\prevdepth=-1000pt \prevdepth0pt\fi
\dimen0=-\prevdepth \advance\dimen0 by20pt\nointerlineskip
\egroup} PACS numbers:~#1}}

\def\@maketitle2{% Puts \@abstract and \@pacs in a {list}.
\@preprint
\@title
\ifdim\prevdepth=-1000pt \prevdepth0pt\fi
\@authoraddress
\@date
\begin{list}{}{\leftmargin=0.10753\textwidth \rightmargin=\leftmargin
\itemsep=1pc\partopsep=-1pc}
\item\@abstract
\item\@pacs
\end{list}
}

\catcode`\@=12

\begin{document}
\draft	
\title{Winding Transitions at Finite Energy and Temperature: An $O(3)$
Model}   
\author{Salman Habib,$^1$ Emil Mottola,$^2$ and Peter Tinyakov$^3$}
\preprint{LA-UR-96-2148}
\address{$^{1}$Theoretical Division, MS B288, Los Alamos National
Laboratory, Los Alamos, NM 87545, USA} 
\address{$^{2}$Theoretical Division, MS B285, Los Alamos National
Laboratory, Los Alamos, NM 87545, USA} 
\address{$^3$Institute for Nuclear Research of the
Russian Academy of Sciences, 60th October Anniversary Prospect, 7a,\\
Moscow, 117312, Russia} 
\date{\today}

\abstract
{Winding number transitions in the two dimensional softly broken $O(3)$ 
nonlinear sigma model are studied at finite energy and temperature. New 
periodic instanton solutions which dominate the semiclassical transition
amplitudes are found analytically at low energies, and numerically for
all energies up to the sphaleron scale. The Euclidean period $\beta$
of these finite energy instantons {\em increases} with energy,
contrary to the behavior found in the abelian Higgs model or simple
one dimensional systems. This results in a {\em sharp crossover} from
instanton dominated tunneling to sphaleron dominated thermal
activation at a certain critical temperature. Since this behavior is
traceable to the soft breaking of conformal invariance by the mass
term in the sigma model, semiclassical winding number transition
amplitudes in the electroweak theory in $3+1$ dimensions should
exhibit a similar sharp crossover. We argue that this is indeed the
case in the standard model for $M_H < 4 M_W$.}

\maketitle2
\narrowtext

%\noindent e-mail: 
%\noindent $^1$habib@predator.lanl.gov\\
%\noindent $^2$emil@pion.lanl.gov\\
%\noindent $^3$peter@amber.inr.free.net 

\section{Introduction}
\label{sec:level1}

Gauge theories of the strong and electroweak interactions are
characterized by a multiple vacuum structure. Tunneling transitions
between different vacua are responsible for physically interesting
effects, such as baryon number violation in the electroweak theory
\cite{tH}.  At zero temperature and energy these winding number
transitions are dominated by the familiar zero energy instanton
solutions of the Euclidean field equations, with vacuum boundary
conditions \cite{BPST}.

At finite temperatures thermal activation over the potential barrier
separating the multiple vacua can take place, in addition to quantum
tunneling. The static classical solution whose energy is equal to the
top of this barrier between neighboring vacua is the sphaleron
\cite{Man}.  At sufficiently high temperatures, transitions between
different winding number sectors are dominated not by quantum
tunneling but by classical thermal activation, with a rate controlled
by the energy of the sphaleron \cite{Kuz}.

In simple one dimensional quantum mechanics, as the temperature is
increased there is typically a smooth crossover from zero energy
quantum tunneling via the instanton to high temperature thermal
activation via the sphaleron \cite{Aff}. The corresponding classical
solutions which interpolate between these two situations are known as
periodic instantons (or ``calorons") \cite{Har,KRT}. These finite
energy solutions of the classical Euler-Lagrange equations possess
turning points at finite Euclidean time $\beta$, and dominate the
semiclassical transition rate at a temperature related to the
Euclidean period by $k_B T=\hbar /\beta$. (Hereafter we set
$\hbar=k_B=c=1$, so that $T = \beta^{-1}$.)

Similar considerations are expected to apply to finite energy winding
number transitions in quantum field theory, although there the
situation is much less well explored and very few of the required
Euclidean periodic classical solutions are known. In particular, it is
not clear {\it a priori} if the smooth crossover of the transition
rate from quantum tunneling to thermal activation is a generic
feature, or if not, on what aspects of the field theory this behavior
depends. Our purpose in this paper is to investigate this question by
finding and studying the finite energy periodic instanton solutions in
a specific model, the $O(3)$ nonlinear sigma model in two dimensions,
modified by a suitable conformal symmetry breaking term. This model is
chosen for the many features it shares in common with spontaneously
broken gauge theories in four dimensions, and in particular, the
standard model of electroweak interactions. In the $SU(2)\times U(1)$
electroweak theory the rate of baryon number violating winding
transitions at high energies remains an open question, despite
considerable efforts in recent years \cite{MMT}. A detailed study of
the solutions of the classical nonlinear field equations of the theory
appears to offer the best hope of addressing this issue
\cite{AGR,RST}.

The $O(3)$ sigma model possesses the advantage of being simple enough
that the zero energy instanton, the finite energy sphaleron, and the
spectrum of linearized fluctuations about each of these solutions are
all known analytically. These analytic results are fixed markers in
the classical solution space that can be used as launching points for
a detailed numerical study of the finite energy periodic instanton
solutions, and which provide useful non-trivial checks on the
numerical methods. The numerical techniques developed in the context
of the sigma model may be applied then with considerably more
confidence to four dimensional gauge theories such as the electroweak
theory. The present work may be regarded as the first step in this
program.

The chief result of our study of periodic instanton solutions in the 
softly broken $O(3)$ sigma model is that the Euclidean period $\beta$
of the finite energy instanton solutions {\em increases} with
increasing energy, {\em i.e.} 
\begin{equation}
{d \beta (E) \over d E} > 0\, ,
\label{sec:dbde}
\end{equation}
for all energies from zero up to the sphaleron.  Since
the Euclidean solution with period $\beta$ is associated with the
winding number transition amplitude at temperature $T=\beta^{-1}$, one
might expect that it should be $T$ which increases with energy, and
not $\beta$. Indeed in simpler models such as the one dimensional
quantum pendulum or the two dimensional abelian Higgs model the very
{\it opposite} behavior to (\ref{sec:dbde}) is found. Our analytic and
numerical study of periodic instanton solutions in the $O(3)$ model
will show that (\ref{sec:dbde}) holds all the way up to the sphaleron
energy, where the periodic instantons merge with the
sphaleron. Continuing further to energies $E > E_{sph}$ results in the
periodic instanton solutions moving off into the complex domain with
still increasing real period.

As we shall see, the main physical consequence of this increase of
period with energy is that there is a sharp crossover from instanton
dominated quantum tunneling to sphaleron dominated thermal activation
at a certain critical temperature (of the order of the mass in the
model), rather than a smooth transition between the two. Since this
feature is traceable to the conformal invariance of the unbroken
$O(3)$ model, the same sharp crossover from quantum tunneling to
thermal activation should be expected in four dimensional gauge
theories where the conformal and spontaneous gauge symmetry breaking
arises from the Higgs sector vacuum expectation value. Thus, the
global picture which emerges from our study of the space of classical
Euclidean solutions and their contribution to fixed energy and fixed
temperature winding number transitions in the $O(3)$ model should be
directly applicable to the electroweak theory. We present evidence
that this is indeed the case, at least for not too large Higgs'
self-coupling, {\em i.e.} for Higgs' mass $M_H < 4 M_W$.

The structure of the paper is as follows. In the next section we
discuss the general properties of periodic instanton solutions and
their contribution to winding number transition rates at finite energy
and temperature, in the context of a quantum mechanical model with
only one degree of freedom, {\em viz.} the simple pendulum. In Section
3 we review the $O(3)$ nonlinear sigma model both before and after
introducing the explicit soft symmetry breaking mass term. The mass
term is necessary in order for a finite energy sphaleron solution to
exist. In Section 4 we consider finite energy instantons in the softly
broken sigma model, employing analytic techniques at low energies
where perturbation theory about the zero energy instantons of the
unbroken model is applicable, and discussing the behavior of the
periodic solutions near the sphaleron energy $E_{sph}$. In Section 5
we describe our numerical techniques and present results for the
finite energy instanton solutions smoothly interpolating between the
two limits. We conclude in Section 6 with a discussion of our results,
and their implications for the analogous winding number transitions in
the electroweak theory.

\section{Periodic Instantons in Quantum Mechanics}
\label{sec:level2}

At finite temperature the partition function for a quantum system may
be expressed as a path integral over configurations with fixed
Euclidean periodicity $\beta$,
\begin{equation}
Z(\beta) = \int_{q(0) = q (\beta)} [{\cal D} q]\  
\exp (-S[q;\beta])
\label{sec:pathint}
\end{equation}
where $S$ is the classical Euclidean action evaluated on the
configuration over one period $\beta$.  If there is a small coupling
in the problem which can be scaled out of the action by $S = s/g^2$
and we consider the weak coupling limit $g^2 \rightarrow 0$, then the
path integral is dominated by the extrema of the classical action,
{\em i.e.} solutions of the classical Euclidean equations of motion
obeying periodic boundary conditions.

Strictly speaking only stable solutions, {\em i.e.} local minima of
the action can contribute to the true equilibrium partition function
at finite temperature. However, we are interested here in calculating
the rate of real time winding number transitions at finite temperature
or energy. In order to contribute to such a transition rate a
classical solution should not be a strict minimum but rather a saddle
point possessing {\em exactly one} negative mode direction.  Analytic
continuation of the semiclassical approximation to $Z(\beta)$ in this
one negative mode direction leads to an imaginary part which may be
interpreted as a real time transition rate \cite{LCC}. A similar
conclusion follows from an analytic extension of the path integral in
(\ref{sec:pathint}) to complex valued configurations \cite{LapMot}.

In systems with only one degree of freedom it is straightforward to
find the relevant periodic instanton solutions with a single negative
mode direction by simple quadrature. Since the energy is an integral
of the motion we have simply to calculate
\begin{equation}
\beta (E) = \oint {dq\over \sqrt{2V(q) - 2E}}
\label{sec:bofE}
\end{equation}
over the periodic trajectory with fixed energy $E$ beginning and
ending at the same turning point of the potential $V(q)$. The
Euclidean action corresponding to this periodic trajectory is
\begin{equation}
S(\beta) = \oint dq\, \sqrt{2V(q) - 2E(\beta)}
\label{sec:Sofb}
\end{equation}
where $E(\beta)$ is obtained by inverting (\ref{sec:bofE}). 

In order to make the discussion definite let us consider the case of a
simple pendulum with the classical periodic potential,
\begin{equation}
V(q) = \omega^2 \left(1-\cos\, q\, \right)\,.
\end{equation}
The periodic local minima of this potential at $q = 2n\pi $ correspond 
to the periodic ground state vacuum structure we find in non-abelian 
gauge theories. The local maxima at $q = (2n + 1)\pi$ are the static 
but unstable ``sphaleron" solutions of this simple model with energy,
\begin{equation}
E_{sph} = 2\omega^2\,.
\end{equation}
In this case the ratio of the zero point energy, $\omega/2$ of the
harmonic potential in the vicinity of the minima to the height of the
potential barrier between minima, {\em i.e.} $\omega/2E_{sph} =
(4\omega)^{-1}$ is the parameter which must be small (in units of
$\hbar = I = 1$) in order to justify a semiclassical treatment of the
quantum pendulum with moment of inertia $I$ about its pivot point.
  
At very low temperatures and energies the winding number transitions
between neighboring minima at $0$ and $2\pi$ are dominated by quantum
tunneling with a rate of order $\exp(-2S_0)$ where $S_0$ is the
classical action for the zero energy kink solution,
\begin{equation}
\cos\left({q_{_0}(\tau)\over 2}\right) =  -\tanh (\omega\tau)
\label{sec:kink}
\end{equation}
which interpolates between the two minima in infinite Euclidean time
$\tau$.  The anti-kink solution is obtained from (\ref{sec:kink}) by
time reversal $\tau \rightarrow -\tau$ and the periodic trajectory
beginning and ending at the same vacuum is a widely separated
kink/anti-kink pair with total action $2S_0 = 16\omega \gg 1$, which
implies that tunneling is exponentially suppressed at zero temperature
or energy.

At the same time it is quite clear that this exponential suppression
disappears at temperatures or energies comparable to $E_{sph}$ when
the pendulum can jump over the barrier between the neighboring minima
by classical thermal activation.  The semiclassical behavior of the
transition rate at intermediate energies and temperatures is easily
found analytically in the case of the pendulum since both of the
integrals (\ref{sec:bofE}) and (\ref{sec:Sofb}) may be expressed in
terms of the complete elliptic integrals $\bf K$ and $\bf E$ as
\begin{equation}
\beta = {4\over\omega}\, {\bf K}(k)
\label{sec:bpend}
\end{equation}
and
\begin{equation}
S = 8\omega\left[ 2{\bf E}(k) - (1-k^2) {\bf K} (k) \right]\,,
\label{sec:Spend}
\end{equation}
where $k$ is the modulus of the elliptic functions, related to the
energy by
\begin{equation}
{d S\over d \beta} = E = 2 \omega^2 (1 - k^2) \qquad {\rm or} \qquad 
k = \sqrt{1 - {E\over 2\omega^2}}\,.
\label{sec:Epend}
\end{equation}
The actual periodic instanton solution $q(\tau)$ with these properties
may be expressed in terms of a Jacobian elliptic function, sn, with
modulus $k$ by
\begin{eqnarray}
&&\cos\left({q(\tau)\over 2}\right) = -k\ {\rm
sn}(\omega\tau,\,k)\nonumber\\ 
&&\rightarrow \left\{\begin{array}{ll}
-\tanh (\omega\tau),& E\rightarrow 0\\
0,& E\rightarrow 2\omega^2 \end{array}\right.
\label{sec:pendinst}
\end{eqnarray}
The corresponding behaviors of the Euclidean period and action in
these limits are 
\begin{equation}
E(\beta) = 2\omega^2 k'^2 \rightarrow \left\{\begin{array}{ll}
0,& \beta\rightarrow \infty\\
E_{sph},& \beta\rightarrow \beta_- \end{array}\right.
\end{equation}
and
\begin{equation}
S(\beta) \rightarrow \left\{\begin{array}{ll}
16\omega,& \beta\rightarrow \infty\\
E_{sph}/T_-,& \beta\rightarrow \beta_- \end{array}\right.
\end{equation}
where 
\begin{equation}
T_- \equiv \beta_-^{-1} = {\omega\over 2\pi}
\end{equation}
and $k'\equiv \sqrt{1-k^2}$ is the complementary modulus of the
elliptic functions.

%PUT FIGURE 1 HERE: PENDULUM ACTION VS. BETA, PLUS
%PERTURBATIVE RESULTS
\vspace{.4cm}
\epsfxsize=7cm
\epsfysize=5.5cm
\centerline{\epsfbox{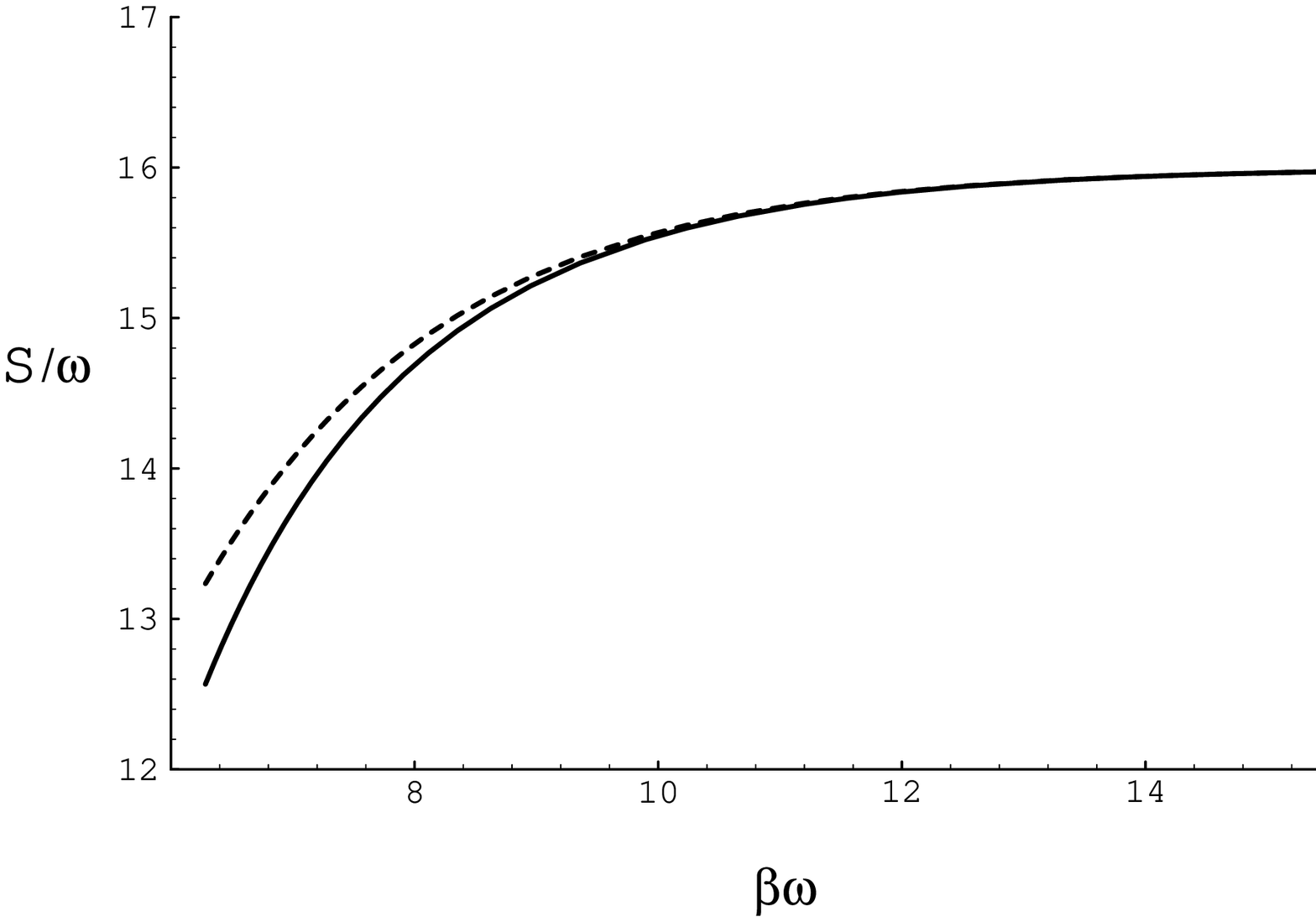}}
\vspace{.35cm}
{FIG. 1. {\small{Exact action, Eqn. (\ref{sec:Spend}), (solid
curve) and perturbative action, Eqn. (\ref{sec:pendSlowE}), (dashed
curve) of the periodic instanton solutions of the simple pendulum
model as a function of their Euclidean period $\beta$. The curves
start at $\beta=\beta_-=2\pi/ \omega$.}}}\\    

In the first limit $E\rightarrow 0$ we recover the vacuum-to-vacuum
kink (or anti-kink) of Eqn. (\ref{sec:kink}) with periodicity $\beta
\rightarrow \infty$, corresponding to a zero temperature winding
transition of the pendulum by quantum tunneling. The second limit, $E
\rightarrow E_{sph}= 2\omega^2$ and $\beta \rightarrow 2\pi/\omega$
corresponds to thermal activation through the unstable, static
sphaleron solution at $q=\pi$ halfway between the two neighboring
periodic vacuum states. The action of the periodic instanton solution
(\ref{sec:pendinst}) interpolates smoothly between these two limits,
as illustrated in Fig. 1.
 
Since $V^{\prime\prime}(q=\pi) = -\omega^2$ the eigenvalues of the
second order fluctuation operator of the Euclidean action with
periodic boundary conditions, expanded around the sphaleron solution
are
\begin{equation}
\lambda_n = \left(2\pi n\over\beta\right)^2 - \omega^2\,.
\label{sec:sphfluc}
\end{equation}
Hence there is clearly exactly one negative mode corresponding to the
unstable $n=0$ perturbation of the sphaleron solution in the high
temperature regime $\beta < \beta_-$, where $\beta_-$ is just the
period of the harmonic oscillation about the inverted potential
$-V(q)$ at $q=\pi$. Thermal activation through the sphaleron dominates
the transition rate for high temperatures $T > T_-$.

For $\beta > \beta_-$, {\em i.e.} low temperatures $T < T_-$, the
sphaleron has more than one negative mode and does not dominate the
semiclassical winding number transition rate. However, in this low
temperature range the periodic instanton given by
Eqn. (\ref{sec:pendinst}) possesses exactly one negative mode by the
following standard argument. Since the periodic instanton obeys the
classical Euclidean Euler-Lagrange equation,
\begin{equation}
{d^2 q\over d\tau^2}  = V'(q)\,,
\label{sec:eom}
\end{equation}
it follows by differentiation that its time derivative is a zero mode
of the second order fluctuation operator, {\em i.e.} 
\begin{equation}
\left(-{d^2 \over d\tau^2} + V''(q)\right) \dot q = 0\,.
\end{equation}
{}From the periodic behavior of $q(\tau)$ it follows that $\dot q$ has
exactly one node in the interval $(0,\beta)$, and hence by a standard
theorem of second order linear differential operators, the fluctuation
operator,
\begin{equation}
-{d^2 \over d\tau^2} + V''(q) = -{d^2 \over d\tau^2} + 2\omega^2 k^2
{\rm sn}^2 (\omega\tau , k) - \omega^2 
\label{sec:flucop}
\end{equation}
must possess exactly one mode with no nodes and a lower ({\em i.e.}
negative) eigenvalue. Hence the finite energy periodic instanton
solution (\ref{sec:pendinst}) dominates the winding number transition
rate for temperatures $T < \omega/2\pi$ and energies $E < 2\omega^2$.

The one negative mode may be understood intuitively as the result of
the attractive interaction between a kink and anti-kink which binds
the pair into a kink/anti-kink ``molecule" and lowers the classical
action of the bound periodic instanton configuration. This attractive
interaction may be studied analytically in perturbation theory at low
energies. Since the same approach proves quite useful in more
complicated models let us review the general method in the pendulum
problem.

Since the kink solution $q_0(\tau)$ of Eqn. (\ref{sec:kink}) has zero
energy we expect the periodic solution for small but finite energy to
consist of an infinite chain of widely separated alternating kinks
and anti-kinks which are very loosely bound. Thus we consider the
trial configuration,
\begin{equation}
q(\tau) = \left\{ \begin{array}{ll}
q_0(\tau)\,, & \qquad 0\le\tau\le{\beta/4}\\
q_0({\beta\over 2} -\tau)\,, & \qquad {\beta/4}\le\tau\le{3\beta/4}\\ 
q_0(\tau - \beta)\,, & \qquad {3\beta/4}\le\tau\le\beta
\end{array}\right. 
\label{sec:trypend}
\end{equation}
defined on the fundamental interval $[0,\beta]$, and consisting of a
kink at $0$ and $\beta$ and an anti-kink halfway between at
$\beta/2$. This configuration may be repeated indefinitely along the
$\tau$ axis to form an infinite chain of alternating kinks and
anti-kinks. The configuration is everywhere continuous and has finite
action per period,
\begin{eqnarray}
S[q(\tau)] &=& \int_0^{\beta} d\tau\, \left[{1\over 2} \dot q^2 +
V(q)\right] \nonumber\\
&=& 4\omega^2 \left[\int_{-\beta/4}^{\beta/4} d\tau\ {\rm
sech}^2 \omega\tau\right.\, \nonumber\\
&& +\, \left. \int_{\beta/4}^{3\beta/4} d\tau\
{\rm sech}^2 \omega\left({\beta\over 2}-\tau\right)\right] \nonumber\\
\qquad &=& 16\omega\, {\rm tanh} \left({\omega\beta\over
4}\right)\nonumber\\ 
\qquad &=& 16\omega - 32\omega e^{-{\omega\beta/2}} + {\cal
O}\left( e^{-\omega\beta}\right) 
\label{sec:actbeta}
\end{eqnarray}
where we retain only the leading $\beta$ dependence for large $\beta$.

The configuration (\ref{sec:trypend}) is also a solution of the
Euclidean equation (\ref{sec:eom}) except at the matching points,
$\tau = \beta/4$ and $\tau = 3\beta/4$ where the first derivative is
discontinuous. That is,
\begin{eqnarray}
&&L[q(\tau)] \equiv - {d^2\over d\tau^2}q + V'(q)\nonumber\\
&& = - 2\delta\left(\tau-{\beta\over 4}\right)\, \dot q_0
\left({\beta\over 4}\right)
+ 2\,\delta\left(\tau-{3\beta\over 4}\right)\, \dot
q_0\left(-{\beta\over 4}\right)\nonumber\\ 
&=&4\omega\, {\rm sech}\left({\omega\beta\over 4}\right)\left[-
\delta\left(\tau-{\beta\over 4}\right) + \delta\left(\tau-{3\beta\over
4}\right)\right] 
\label{sec:linpend}
\end{eqnarray}
These delta functions act as point sources for the field $q(\tau)$.
If we expand about the trial configuration (\ref{sec:trypend}), $q
\rightarrow q + \xi$, then the non-zero result of
Eqn. (\ref{sec:linpend}) implies that there are linear source terms for
the fluctuations $\xi$.  The effect of these sources on the action
functional may be found by shifting the fluctuation
\begin{equation}
\xi(\tau) \rightarrow \xi(\tau) - \int_0^{\beta} d\tau'\,
G_{\beta}(\tau,\tau')\, L[q(\tau')]
\label{sec:shft}   
\end{equation}
to remove the linear term \cite{LapMot,Boga}. The periodic Green's
function $G_{\beta}(\tau,\tau')$ obeys
\begin{eqnarray}
\left[ - {d^2\over d\tau^2} + V''(q)\right]G_{\beta}(\tau, \tau') &=& 
\delta_{\beta} (\tau-\tau') \nonumber\\
= {1\over\beta} \sum_{n=-\infty}^{\infty}
e^{2\pi i(\tau-\tau')n/\beta}\,.
\end{eqnarray}
To leading order in the exponentially small tail
$e^{-{\omega\beta/2}}$ we can replace $V''(q)$ by its vacuum value,
$\omega^2$ and use the free (perturbative) periodic Green's function,
\begin{equation}
G_{\beta}(\tau, \tau') = {1\over 2\omega}{1\over \sinh
\left({\beta\omega / 2}\right)} \cosh \omega\left(\vert
\tau-\tau'\vert - {\beta\over2}\right) 
\label{sec:therG}
\end{equation}
for $\tau$ and $\tau'$ in the fundamental interval $[0,\beta]$.

To leading order in $L[q]$ the effect of the shift (\ref{sec:shft}) to
remove the linear term is to alter the action at quadratic order, {\em
i.e.}
\begin{eqnarray}
&&S[q(\tau)] \rightarrow S[q(\tau)]\nonumber\\
&& - {1\over 2}\int_0^{\beta} d\tau\,
\int_0^{\beta} d\tau'\, L[q(\tau)]\, G_{\beta}(\tau,\tau')\,
L[q(\tau')]\,.   
\end{eqnarray}
Substituting (\ref{sec:linpend}) and (\ref{sec:therG}) into this extra
term leads to the result that the action of the periodic instanton is
the action of the trial kink/anti-kink configuration
(\ref{sec:actbeta}) shifted by
\begin{eqnarray}
& &-8\omega^2 {\rm sech}^2\left({\omega\beta\over 4}\right)\left[
G_{\beta}\left({\beta\over 4}, 
{\beta\over 4}\right) + G_{\beta}\left({3\beta\over 4}, {3\beta\over
4}\right) 
\right.\nonumber\\
& &\left. - G_{\beta}\left({\beta\over 4},{3\beta\over 4}\right) -
G_{\beta}\left({3\beta\over 4},{\beta\over 4}\right)\right]\nonumber\\ 
&=& -32\omega\,e^{-{\omega\beta/2}} +{\cal
O}\left(e^{-\omega\beta}\right)\,. 
\end{eqnarray} 
Adding this shift to (\ref{sec:actbeta}) we find that the action of the
periodic instanton in the pendulum model is
\begin{equation}
S(\beta) = 16\omega - 64\,\omega\, e^{-{\omega\beta/2}} + {\cal
O}\left( e^{-\omega\beta}\right) 
\label{sec:pendSlowE}
\end{equation}
while its energy becomes
\begin{equation}
E(\beta) = {d S\over d\beta} = 32\omega^2 e^{-{\omega\beta/2}} +
{\cal O}\left( e^{-\omega\beta}\right) 
\label{sec:pendElowE}
\end{equation}
to leading order in the expansion in powers of $E/E_{sph} =
16\exp(-{\omega\beta/2})$. The negative second term in
Eqn. (\ref{sec:pendSlowE}) is the effect of the short-ranged attractive
interaction between neighboring kinks and anti-kinks along the chain
that lowers the action functional to first order in the two-body
interaction $\exp(-\omega (|\tau_1 - \tau_2|))$ between them.

This result and this general method of patching together zero energy
kinks and anti-kinks can be checked in the pendulum example by
comparison to the exact results (\ref{sec:bpend}-\ref{sec:Epend}) in
terms of the elliptic functions. Indeed in the low energy limit the
modulus $k\rightarrow 1$ and the expansion of the complete elliptic
integrals in powers of the complementary modulus,
\begin{equation}
k'\equiv \sqrt{1-k^2} \rightarrow 4\, e^{-{\omega\beta/4}}
\end{equation}
yields
\begin{eqnarray}
S(\beta) &=& 16\,\omega\left( 1 - {k'^2\over 4} + {\cal
O}(k'^4)\right)\nonumber\\  
&=& 16\,\omega - 64\, \omega\, e^{-{\omega\beta / 2}} + {\cal
O}\left( e^{-\omega\beta}\right) 
\end{eqnarray}
and
\begin{equation}
E(\beta) = 2\omega^2 k'^2 = 32\, \omega^2\, e^{-{\omega\beta / 2}}
+ {\cal O}\left( e^{-\omega\beta}\right) 
\end{equation}
which coincides with (\ref{sec:pendSlowE}) and
(\ref{sec:pendElowE}). The exact and perturbative results for the
action as a function of $\beta$ are compared in Fig. 1.

%PUT EXACT ACTION OVER BIG RANGE/DASHED CURVE PLUS STRAIGHT LINE
%HERE.
\vspace{.4cm} 
\epsfxsize=7cm
\epsfysize=5.5cm
\centerline{\epsfbox{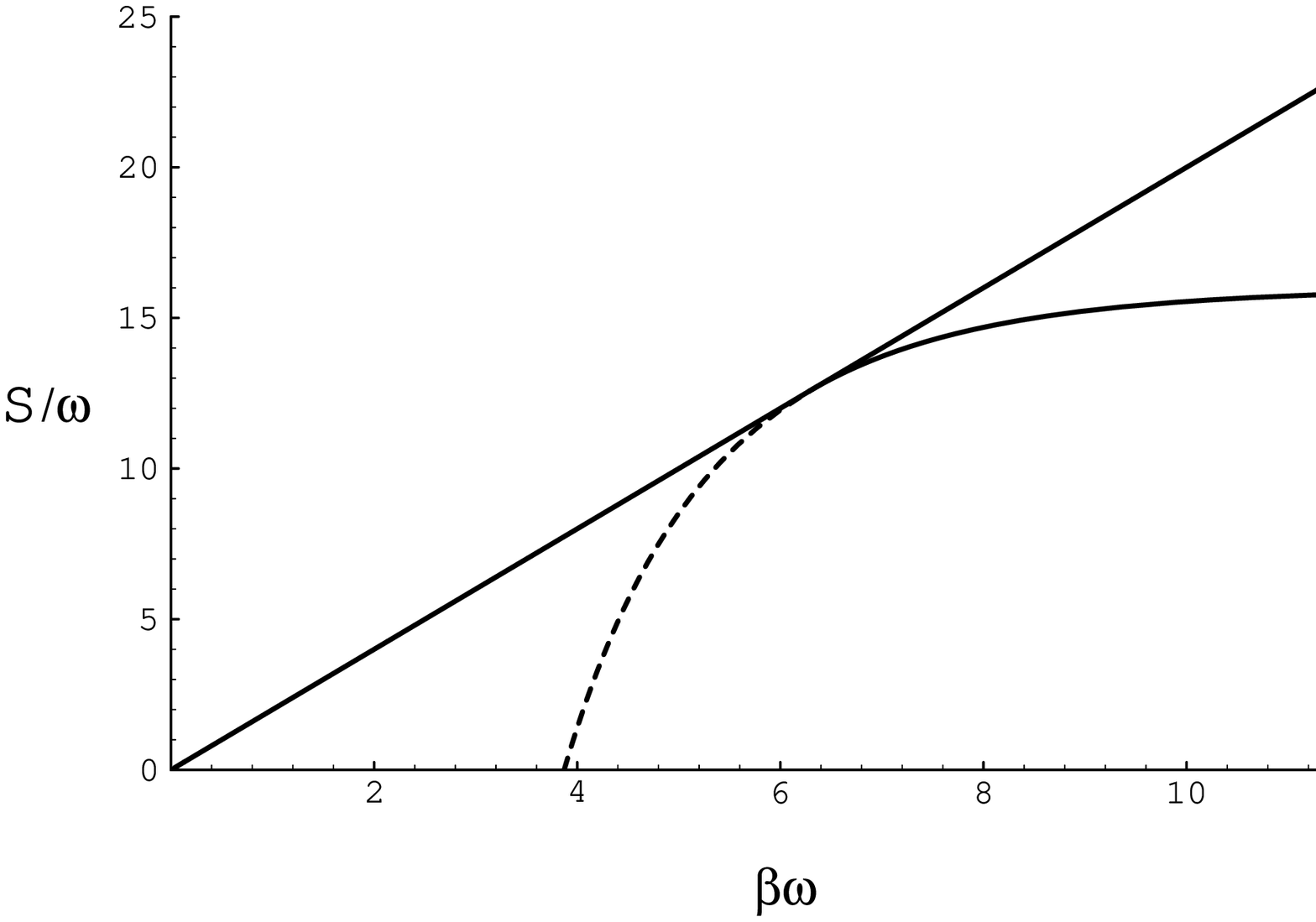}}
\vspace{.35cm}
{FIG. 2. {\small{The action of the periodic instanton and
sphaleron as a function of $\beta$ for the simple pendulum model.  The
instanton solution goes from real (solid curve) to complex (dashed
curve) at $\beta=\beta_-$, where the sphaleron action $S=E_{sph}\beta$
(straight line) is tangent to the curve. At $\beta =
\beta_{cr}=8/\omega > \beta_-$ the zero energy instanton action $2S_0$
equals the sphaleron action.}}}\\   

As $\beta$ is decreased from infinity to $\beta_-$ and $E$ is
increased from zero to $E_{sph}$, the kink/anti-kink molecule becomes
more tightly bound and the nonlinear interactions between the kink and
anti-kink become more important. The action of the periodic instanton
(\ref{sec:pendinst}) decreases from $2S_0 = 16\omega$ to $S(\beta_-) =
E_{sph}\beta_- = 4\pi \omega < 2S_0$ and it is easy to see from the
properties of the elliptic functions involved that the derivative,
$d\beta(E)/dE$ is {\em negative} everywhere in the interval $(0,
E_{sph})$. At $E=E_{sph}$, $\beta = \beta_-$, $S = E_{sph}\beta_-$,
and the periodic instanton solution (\ref{sec:pendinst}) becomes {\em
identical} to the sphaleron. Beyond this point we no longer have a
real periodic instanton solution. This presents no difficulty as
analytic continuation of the solution (\ref{sec:pendinst}) for $E >
E_{sph}$ is easily accomplished by allowing the modulus $k$ to become
purely imaginary. Then the solution $q(\tau)$ becomes complex (in
fact, $\pi$ plus a purely imaginary function of $\tau$), but the
period (\ref{sec:bpend}) and action (\ref{sec:Spend}) remain real and
continue to decrease as the energy is increased. However, the
fluctuation operator (\ref{sec:flucop}) now has additional negative
modes and hence the analytically continued solution no longer
dominates the winding number transition rate, its place having been
taken by the static sphaleron for $T > \omega/2\pi$. At $\beta =
0.61627 \beta_-$ and $E = 5.74914 E_{sph}$ the action
(\ref{sec:Spend}) of the complex instanton actually plunges through
zero and then becomes negative, finally going to $-\infty$ as
$E\rightarrow\infty$ and $\beta \rightarrow 0$. The situation is most
conveniently illustrated graphically in Fig. 2. Notice that, in this
simple pendulum model the zero temperature tunneling exponent $2S_0$
is equal to the thermal activation Boltzmann exponent $E_{sph}/T$ at
$T = \omega/8$ which is {\em less than} $T_- = \omega/2\pi$.
 
%PUT THE MEXICAN HAT AND OTHER DIRECTION FIGURES HERE (FIGS 3a and 3b).
\vspace{.4cm} 
\epsfxsize=6cm
\epsfysize=4cm
\centerline{\epsfbox{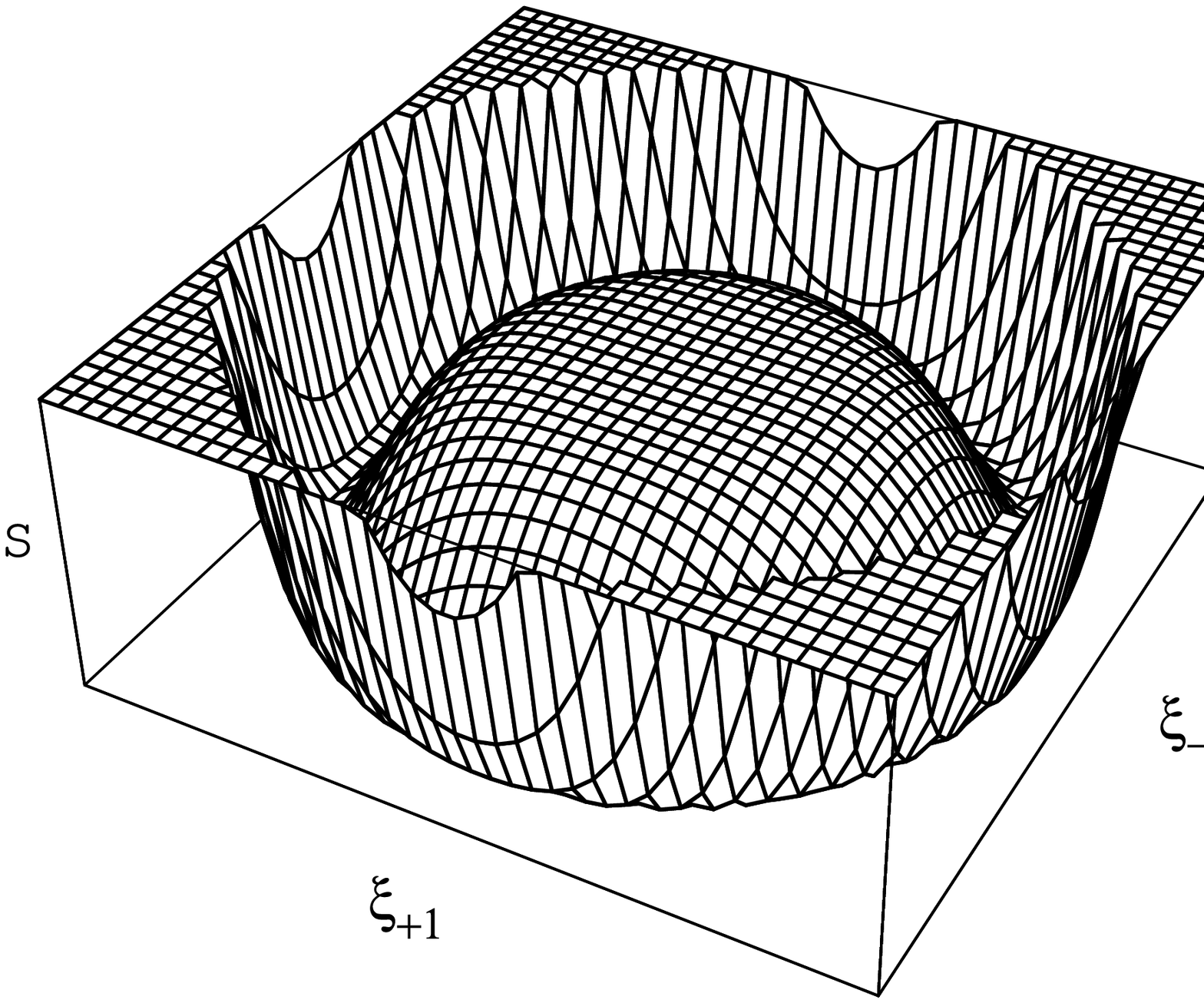}}
\vspace{.35cm}
\epsfxsize=6cm
\epsfysize=4cm
\centerline{\epsfbox{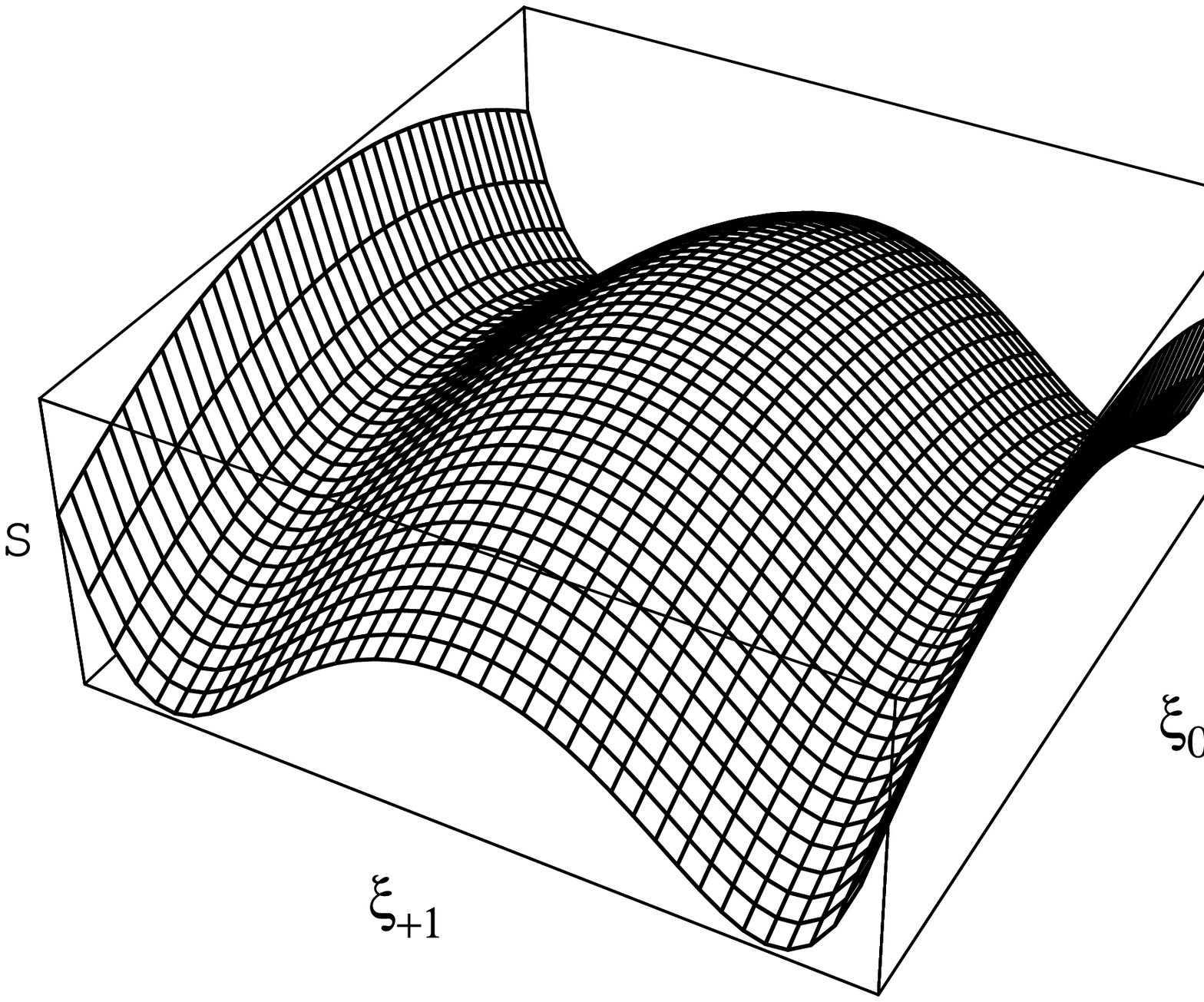}}
\vspace{.35cm}
{FIGS. 3. {\small{The action $S$ as a function of the
variables $\xi_{\pm 1}$ (a), corresponding to the two near zero mode
directions, and as a function of one of these variables and the
negative mode direction $\xi_0$ (b), for fixed $\beta$ slightly larger
than $\beta_-$.  The ring of extrema of $S$ in the first figure (due
to the arbitrary phase in Eqn. \ref{sec:epend}) or the two outer extrema
in the second figure correspond to the periodic instanton solution. As
$\beta$ approaches $\beta_-$ these extrema merge with the sphaleron in
the center. Since the periodic instanton solution has one negative
mode and lower action than the sphaleron, it dominates the winding
number transition rate which is controlled by the lowest available 
saddle in (b).}}}\\

The merging and resplitting off into the complex domain of the
periodic instanton solution as $\beta$ is decreased through $\beta_-$
is an interesting aspect of the behavior we have just sketched. This
may be understood from the fact that near $\beta =\beta_-$ the
eigenvalues of the static sphaleron fluctuation operator,
$\lambda_{\pm 1}$ in (\ref{sec:sphfluc}), go through zero.  Whenever a
zero mode appears in the second order fluctuation operator this
indicates the existence of a nearby solution of the classical
equations.  When the zero mode is not related to a symmetry of the
action but instead appears only at special values of a parameter in
the boundary conditions, then the solution generally splits off or
{\em bifurcates} into two (or more) different solutions as the
parameter is varied. Conversely, as the parameter is varied in the
opposite direction two or more solutions {\em merge} at the critical
value of the parameter.  This is just the case for the static
sphaleron solution as the parameter $\beta$ is varied through
$\beta_-$.  At $\beta_-$ we find another classical solution with the
same Euclidean period splitting off or merging with the static
sphaleron in the (generally complex) direction of the zero mode in
function space. Hence near $\beta = \beta_-$ the periodic instanton
solution is given approximately by a small perturbation of the
sphaleron in the relevant zero mode direction, {\em viz.}
\begin{equation}
q(\tau) \simeq q_{sph} + \delta\, \cos (\omega \tau + \phi)
\label{sec:epend}
\end{equation}
where $\phi$ is an arbitary phase of the periodic solution and
$\delta$ is a small parameter that goes to zero as
$\beta\rightarrow\beta_-$.  The arbitary phase informs us that in fact
there is a one parameter $U(1)$ family of periodic instanton solutions
which merge and then resplit off from the sphaleron at $\beta_-$ which
accounts for the two zero mode directions, $n= \pm 1$ at $\beta
=\beta_-$.

This situation is also easier to visualize graphically. Let us
suppress all but the relevant zero mode directions and represent the
action functional $S$ as a function only of a few 
variables $\xi_i$ together with the parameter $\beta$.  Clearly as
$\beta$ varies the surfaces of $S= const.$ change, as do the location
of the extrema, $\partial S/\partial \xi_i = 0$.  At certain values of
$\beta$ if two of the extrema split off or merge, there is a zero
eigenvector of the second order fluctuation matrix $\partial^2
S/\partial \xi_i\partial \xi_j$. Since the number of extrema is
determined by the global properties of $S$, generally the bifurcation
of extrema does not mean that their total number changes
discontinuously, but rather that real extrema move into the complex
domain as the parameter $\beta$ is varied through $\beta_-$. The
$U(1)$ invariance of the action under the arbitrary phase $\phi$
implies that a ring of equal action extrema converges on the sphaleron
and then becomes complex for $\beta < \beta_-$. This action as a
function of the three relevant variables $\xi_{\pm 1}$ and $\xi_0$ is
illustrated in Figs. 3.

To summarize this discussion, the physical interpretation of the
properties of the classical Euclidean periodic solutions for winding
number transitions in the simple pendulum model is clear. At low
temperatures $T < 1/\beta_-$ the periodic instanton solution dominates
the vacuum-to-vacuum winding number transition amplitude which is
exponentially suppressed. As the temperature is raised the classical
periodic solutions become a more closely spaced chain of kinks and
anti-kinks with a larger nonlinear interaction between them that
lowers the action. The tunneling rate becomes less suppressed as the
turning points of the periodic solution move inwards toward their
midpoint at $q=\pi$, corresponding to less quantum tunneling and more
thermal motion in the allowed region. At $T=T_- = \omega/2\pi$ the
turning points become identical with the sphaleron at $q=\pi$ and
tunneling has been replaced by purely classical thermal activation
which is still only slightly less exponentially suppressed than the
zero temperature transition rate. At this and all higher temperatures
the winding number transition rate is dominated by the sphaleron which
has exactly one negative mode for $T>T_-$. The periodic instanton
solution moves off into the complex domain and no longer contributes
to the transition rate, which becomes less and less exponentially
suppressed as the temperature is raised further. Finally at
temperatures greater than of order $E_{sph}= 2\omega^2$ the
exponential suppression disappears entirely and the pendulum swings
freely around its pivot, the probability of finding the pendulum
anywhere then becoming nearly uniform as the temperature is raised
still further.
     
This fixed temperature discussion has a natural analog at fixed
energy. The probability of making a winding number transition at fixed
energy between $E$ and $E + dE$ may be expressed in the form,
\begin{equation}
P(E)dE= \sum_{i,f} |\langle f\vert{\cal S} {\cal P}_E\vert i\rangle|^2
\; , 
\label{sec:P[E]} 
\end{equation}
where ${\cal S}$ is the ${\cal S}$-matrix, ${\cal P}_E$ is a projector
onto energy $E$ and the initial and final states $\vert i\rangle$ and
$\vert f\rangle$ lie in different winding sectors in the energy
interval $E$ to $E + dE$. Periodic instantons appear again as the
configurations which saturate $P(E)$, and the probability is given,
with exponential accuracy, by the exact analog of the
quantum-mechanical formula, \cite{Aff,KRT}
\begin{equation}
P(E)dE\sim \exp(-W(E))dE \,, \label{sec:P[E]semicl}  
\end{equation}
where
\begin{equation}
W(E) = S(\beta(E)) - E\beta(E)
\end{equation}
is the Legendre transform of $S(\beta)$. The relevant instanton
solutions should be considered now as fixed in energy rather than in
periodicity $\beta$. 
  
The initial and final multi-particle states can be read off from the
analytic continuation of the finite energy periodic instanton into
Minkowski time at its turning points. Hence periodic instanton
solutions to the classical Euclidean equations contain non-trivial
information about multi-particle transition amplitudes between
different winding number sectors at finite energy. It has been
suggested that by suitably modifying the boundary conditions in the
complex time plane, information about transition amplitudes from few
particle initial states to many particle final states may be obtained
as well \cite{RST}.

The fixed temperature transition rate can be reconstructed from the
fixed energy tunneling probability by weighting with a Boltzmann
distribution and integrating over energy. To exponential accuracy we
have
\begin{eqnarray}
\Gamma (T)&=&\int_0^{\infty} dE\,\exp\left(- {E\over T}\right) P(E)
\nonumber\\ 
&\sim& \int_0^{\infty} dE\,\exp\left(- {E\over T}- W(E)\right)~.
\label{sec:rate}
\end{eqnarray}
Evaluating this integral by the method
of steepest descent (which is valid in the limit of arbitrarily weak
coupling $\omega^{-1} \sim g^2\rightarrow 0$, provided $T$ is coupling
independent) gives the saddle point condition,
\begin{equation}
{1\over T} = -{dW\over dE} = \beta (E)\,,    \label{sec:saddle}
\end{equation}
by the properties of the Legendre transform $W = S - E\beta$.  Hence
we recover the connection between the temperature and Euclidean
periodicity of the instanton solution in this way.  Evaluating the
second derivative of the exponent in (\ref{sec:rate}) at this saddle
point yields
\begin{equation}
-{d^2 W \over dE^2} = +{d\beta (E)\over dE}\,,
\label{sec:secder}
\end{equation}
which tells us that only solutions for which the quantity in
(\ref{sec:secder}) is {\em negative} can contribute to the fixed
temperature rate $\Gamma(T)$. On the other hand, classical solutions
for which this derivative is positive cannot contribute to the
integral (\ref{sec:rate}) since they are local {\em minima} rather
than maxima of the the exponent in the integrand. In the case of the
simple quantum pendulum example considered in detail in this section
(\ref{sec:secder}) is indeed negative and the periodic instantons
given by (\ref{sec:pendinst}) do contribute to the finite temperature
transition rate $\Gamma (T)$ in the expected way, for all temperatures
$T \le T_-$. We turn now to the $O(3)$ sigma model where the behavior
of the periodic instanton solutions, the corresponding quantity
(\ref{sec:secder}) and the crossover from quantum tunneling to thermal
activation are all quite different.

\section{The $O(3)$ Sigma Model}
\label{sec:level3}

The two-dimensional $O(3)$ nonlinear sigma model is defined by
the Euclidean action functional, 
\begin{equation}
S={1\over 2g^2} \int dx\,d\tau \,(\partial_{\mu}n^a)^2\; ,
\label{sec:action0}
\end{equation}
where $n^a(x)$, $a=1,2,3$ are three components of a unit vector,
$n^an^a=1$, and $\mu = \tau, x$. The constraint on the magnitude of
$n^a$ at every spacetime point is the source of nonlinearity in the
model. A convenient parameterization in which the constraint is
removed at the price of explicit nonlinearity is the complex field
definition,
\begin{equation}
w \equiv {n_1 + i n_2 \over 1 - n_3}\,.
\end{equation}
Introducing the complex coordinate
\begin{equation}
z \equiv x + i\tau \equiv r e^{i\theta}
\end{equation}
in terms of the spatial position $x$ and Euclidean time $\tau$ enables
us to rewrite the action (\ref{sec:action0}) in the equivalent forms,
\begin{eqnarray}
S &=&{2\over g^2} \int dx\,d\tau\, {\partial_{\mu}w\partial_{\mu}\bar
w \over (1 + \bar w w)^2}\qquad {\rm or}\nonumber\\  
S &=&{4\over g^2} \int dx\,d\tau \, {1 \over (1 + \bar w w)^2} 
\left({\partial w \over \partial z}{\partial \bar w \over \partial
\bar z} + {\partial \bar w \over \partial z}{\partial w \over \partial
\bar z}\right)\,,  
\label{sec:actionc}
\end{eqnarray}
where the overbar denotes complex conjugation and $\bar w =\overline{
w (z, \bar z)} =w (\bar z, z)$. 

This model possesses some remarkable similarities with pure gauge
theories in four dimensions. The most important properties which
concern us here are the conformal invariance of the classical action
$S$, and the existence of a periodic vacuum structure in the model. In
connection with the conformal invariance we observe that the coupling
constant $g$ is dimensionless and there is no length scale in
$S$. Correspondingly, the quantum theory is logarithmically
renormalizable and in fact, asymptotically free in the coupling $g$
\cite{Poly}.

The periodic structure should be clear from the fact that an $O(3)$
rotation by $2\pi$ around any axis brings the vector $n^a$ back to
itself. The topological winding number associated with this field
periodicity will be made explicit if we identify the points at
infinity of the complex plane in the coordinate $z=x + i\tau$. Then
the plane has the topology of the sphere $S^2$. Since the field $n^a$
is also constrained to lie on $S^2$, the $n^a$ field is a map from
$S^2$ to $S^2$ which is characterized by an integer winding number,
given explicitly by
\begin{eqnarray}
Q &=& {1\over 8\pi}\int dx\,d\tau \ \epsilon^{\mu\nu}\epsilon_{abc} 
n^a \partial_{\mu} n^b \partial_{\nu} n^c \nonumber\\
&=& {1\over \pi} \int dx\,d\tau \ {1 \over (1 + \bar w w)^2} 
\left({\partial w \over \partial z}{\partial \bar w \over \partial
\bar z} - {\partial \bar w \over \partial z}{\partial w \over \partial
\bar z}\right)\,. 
\end{eqnarray}
Comparing the latter form of the topological winding number with the
action (\ref{sec:actionc}) it is clear that the action in any integer
$Q$ topological sector is bounded from below, {\em i.e.}
\begin{equation}
S \ge {4\pi\over g^2} \ \vert Q\vert
\end{equation}
and moreover that this bound is saturated by meromorphic functions $w$
of the complex variable $z$ or $\bar z$ \cite{sigmainst}. In
particular the conformal map of $S^2$ to $S^2$, topologically
equivalent to the identity map is the one instanton solution of the
Euclidean equations with $Q=1$. This solution can be written in the form,
\begin{equation}
w_0 (z) = {\rho\over z} = {\rho\over r}e^{-i\theta}
\label{sec:inst}
\end{equation}
and has action 
\begin{equation} 
S_0 = {4\pi\over g^2}\,.
\end{equation}
The anti-instanton solution is obtained from (\ref{sec:inst}) by
Euclidean time reversal which in the $w$ description is equivalent to
complex conjugation.

The instanton and anti-instanton solution have zero Euclidean energy,
\begin{equation}
E_0 = {1\over 2g^2}\int dx \left(- \partial_{\tau}n^a
\partial_{\tau}n^a + \partial_x n^a \partial_x n^a\right) = 0\,.
\label{sec:ener}
\end{equation}
They correspond respectively to the classical Euclidean path of least
action and its time reverse, connecting periodic vacua separated by
unit winding number. Together the instanton and anti-instanton
constitute a periodic trajectory beginning and ending at the same
vacuum state, with one negative mode corresponding to the attractive
interaction between them, just as in the simple pendulum model.  Hence
in the limit of weak coupling $g^2 \rightarrow 0$ the rate for a unit
winding number vacuum to vacuum tunneling transition is $\exp(-2S_0) =
\exp (-8\pi/g^2)$ to exponential accuracy, and is strongly suppressed.

Because of the conformal invariance of the action (\ref{sec:action0}),
the zero energy instanton can have arbitrary scale $\rho$. This
implies that the potential energy barrier between winding number
sectors, {\em i.e.} the second term of (\ref{sec:ener}) evaluated at
$\tau = 0$ which is proportional to $\rho^{-1}$, can be made
arbitrarily small and no finite energy sphaleron solution exists in
the symmetric $O(3)$ model. This feature is quite different from the
simple pendulum and abelian Higgs models but is shared by the pure
Yang-Mills theory in four dimensions.
 
In the electroweak theory conformal invariance is broken in the Higgs
sector which then makes possible the existence of a classical
sphaleron solution. In the $O(3)$ sigma model the conformal invariance
may be broken by adding to the action (\ref{sec:action0}) the explicit
mass term
\begin{equation}
S_m={m^2\over g^2}\int dx\,d\tau\ (1+n_3)\; ,
\label{sec:action1}
\end{equation}
which also violates the $O(3)$ symmetry and fixes the vacuum state to
be $n^a_{(vac)}=(0,0,-1)$. In the Heisenberg spin language this
corresponds to placing the system in an external magnetic field which
aligns all the spins in the direction $n^a_{(vac)}$ at zero
temperature.

When the sigma model has been modified in this way by the introduction
of the soft symmetry breaking mass term $S_m$, 
\begin{equation}
S \rightarrow S + S_m\,,
\label{sec:Stot}
\end{equation}
the instanton configuration with arbitary non-zero $\rho$ in
(\ref{sec:inst}) ceases to be an exact solution of the field
equations, but there is now an exact sphaleron solution
\cite{Mottola&Wipf,Fun}, namely,
\begin{eqnarray}
n_{(sph)}^1(x)&=&-2\tanh (mx)\, {\rm sech}(mx)\,,\nonumber\\
n_{(sph)}^2(x)&=&0\,,\nonumber\\
n_{(sph)}^3(x)&=&-1+2\,{\rm sech}^2(mx)
\label{sec:sphaleron} 
\end{eqnarray}
with finite energy, 
\begin{equation}
E_{sph}={8m\over g^2}\,. \label{sec:sphener}
\end{equation}
Geometrically, this solution maps the infinite spatial line onto a
great circle beginning and ending at the south pole, $n^3_{(vac)} =
-1$.  The sphaleron lies exactly halfway between the vacua of winding
number $0$ and $1$, and $E_{sph}$ is therefore the height of the
potential energy barrier between these vacuum states, which is now
fixed and finite.

Let us recall that the spectrum of time independent perturbations
around the static sphaleron has exactly one normalizable negative mode
with the eigenfunction $u^a_-(x)=(0,\,{\rm sech}^2(mx),\,0)$ and the
eigenvalue $\epsilon_-^2=-3m^2$.  There are two zero modes of the
spatial fluctuation operator corresponding to spatial translations of
the sphaleron position and rotation of its great circle trajectory
around the $n^3$ axis. The remaining eigenvalues are strictly positive
and form a continuous spectrum with $\epsilon_+^2 \ge m^2$. Taking the
periodic $\tau$ dependence of the fluctuations into account yields the
following eigenvalues for the full second order fluctuation operator
of the action around the sphaleron solution,
\begin{equation}
\lambda_{n,\epsilon}= \left({2\pi n\over\beta}\right)^2 + \epsilon^2\,.
\label{sec:eigen}
\end{equation}
Periodic boundary conditions are the appropriate ones for
contributions to the path integral at finite temperature $T
=\beta^{-1}$. By the same argument as in the one dimensional pendulum
the existence of one and only one negative eigenvalue (\ref{sec:eigen}),
namely $n=0$ and $\epsilon^2 = \epsilon^2_- = -3m^2$ for $\beta <
\beta_-\equiv 2 \pi /|\epsilon_-| = 2\pi/(\sqrt{3}m)= 3.628\,m^{-1}$
implies that the sphaleron configuration contributes to the winding
number transition rate for temperatures $T > T_-$. The sphaleron
contribution to the rate per unit one dimensional spatial volume can
be expressed in the notation of the present paper in the form, 
\cite{Mottola&Wipf}
\begin{equation}
\Gamma_{sph} = {2\over \pi g^2} {m T\over
\sin\left({\vert\epsilon_-\vert / 2T}\right)} \exp\left[ -
E_{sph}/T - h (m/T)\right]\,, 
\label{sec:sphrate}
\end{equation} 
where
\begin{eqnarray}
&& h(m/T) \equiv \nonumber\\
&& -{4m\over \pi}\int_0^{\infty}dk\,\left[ {1\over
\omega_k^2} + {1\over \omega_k^2 + 3m^2}\right]\ln \left( 1 -
e^{-\omega_k/T}\right)\,, 
\end{eqnarray}
and
\begin{equation}
\omega_k \equiv \sqrt{k^2 + m^2}\,.
\end{equation}
The asymptotic form of this function in the high temperature
limit $m\beta \ll 1$ is
\begin{equation}
h(m/T) \rightarrow -3\ln (m/T) - {4m\over \pi T}\ln (m/T) - C
+ {\cal O} (m/T)\,,
\end{equation}
where we take this opportunity to correct an error in the numerical
evaluation of the constant $C$ in Eqn. (5.7) of
Ref. \cite{Mottola&Wipf}, {\em viz.},
\cite{GR}
\begin{eqnarray}
C &=& {2\over \pi} \int_0^{\infty} dx \left[{1\over x^2 + 1} + {1\over
x^2 + 4} \right] \ln (x^2 + 1)\nonumber\\ 
&=& \ln\, 12 = 2.4849\ .
\end{eqnarray}
As a result, the constant defined in Eqn. (5.17) of
Ref. \cite{Mottola&Wipf} changes as well, becoming $K = 2e^C/\sqrt 3 =
8\sqrt 3 = 13.8564\ $.  

The sphaleron thermal activation rate (\ref{sec:sphrate}) is less than
the instanton mediated tunneling rate unless $T > T_{cr}$ where the
crossover temperature,
\begin{equation}
T_{cr} = {m\over \pi} >  T_- = {\sqrt 3 \over 2}{m\over \pi}
\label{sec:Tcr}
\end{equation}
is defined by equating the quantum tunneling and thermal activation
exponents, $2S_0=E_{sph}/T_{cr}$. Hence although the sphaleron
provides a second semiclassical escape path for winding number
transitions for all temperatures above $T_-$, its contribution is
exponentially subdominant until the temperature $T\ge T_{cr} > T_-$
where there is a {\em sharp crossover} from quantum tunneling mediated
by the zero scale singular instanton to thermal activation via the
static sphaleron, in the semiclassical weak coupling approximation
$g^2 \ll 1$.

The fact that the crossover between winding number transitions
dominated by quantum tunneling and those dominated by thermal
activation takes place at a temperature $T_{cr}$ greater than that at
which the $n=1$ eigenvalue (\ref{sec:eigen}) goes through zero is a
qualitative difference from the situation in one dimensional quantum
mechanics or the abelian Higgs model in $1 +1$ dimensions. In each of
these examples the temperature at which the sphaleron and
instanton/anti-instanton exponents are equal is {\em smaller} than the
corresponding $T_-$ of the sphaleron negative mode. In such cases the
crossover between zero temperature quantum tunneling and finite
temperature thermal activation is smooth and the sphaleron dominates
the transition rate for all $T \ge T_-$. The situation for the $O(3)$
sigma model is illustrated in Fig. 4 where the numerical results from
Section 5 are incorporated. 

%FIG4 GOES HERE (NEWACTCOMP)
\vspace{.4cm} 
\epsfxsize=6cm
\epsfysize=5cm
\centerline{\epsfbox{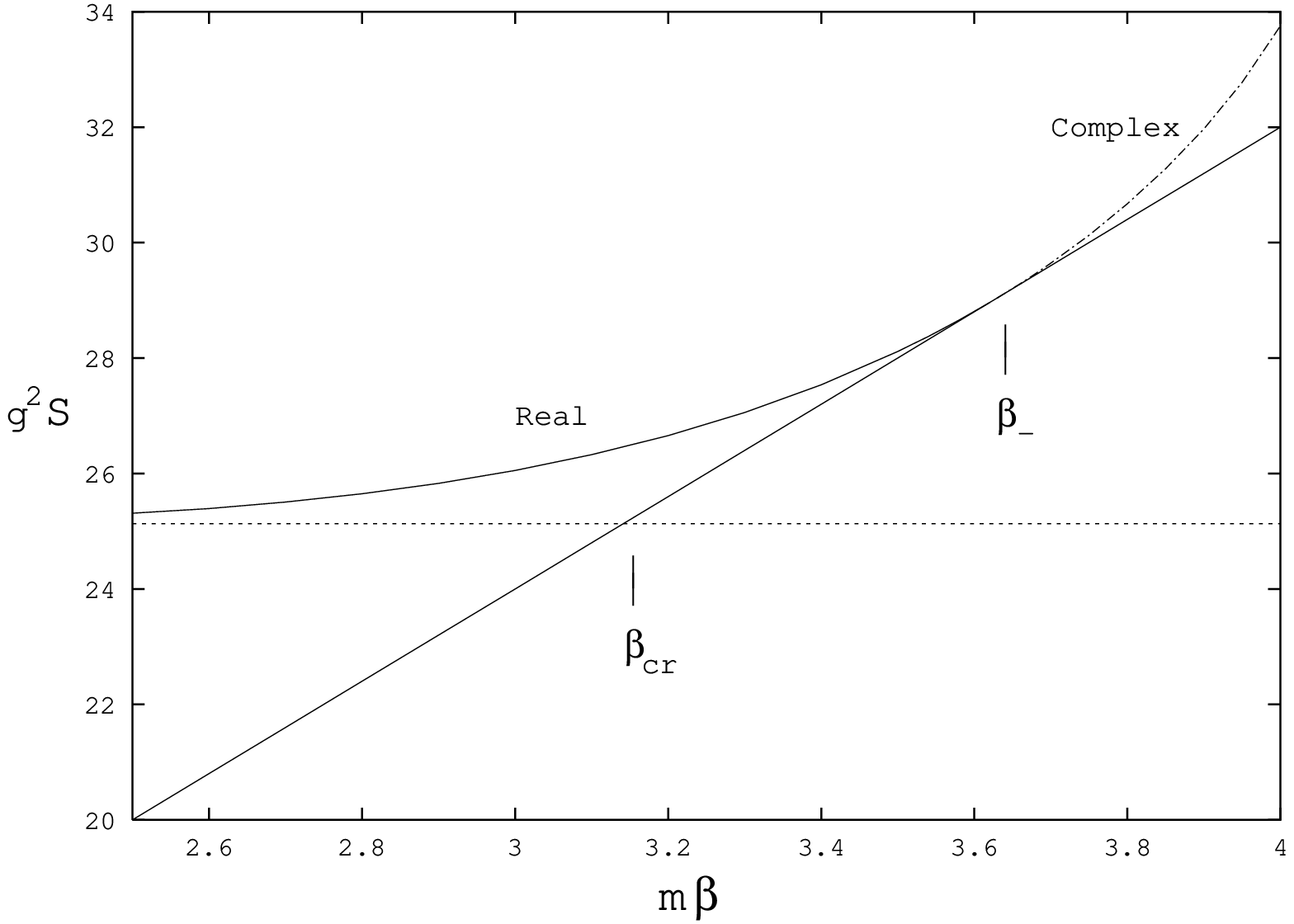}}
\vspace{.35cm}
{FIG. 4. {\small{The action of the periodic instanton and
sphaleron as a function of $\beta$ for the $O(3)$ sigma model. As in
Fig. 2 the instanton solution goes from real (solid curve) to complex
(dot-dashed curve) at $\beta=\beta_-$, where the sphaleron action
$S=E_{sph}\beta$ is tangent to the curve. However, in contrast to
Fig. 2 the value of $\beta_{cr}$ at which the sphaleron action equals
the zero energy action $2S_0$ (dashed horizontal line) is less than
$\beta_-$.}}}\\

As is easy to see from scaling arguments (Derrick's theorem) the mass
term $S_m$ now drives the instanton size $\rho \rightarrow 0$, so
there is no exact instanton of the form (\ref{sec:inst}) except the 
singular one with $\rho=0$. However, for small but finite $\rho$ the 
instanton configuration
(\ref{sec:inst}) remains an approximate solution for $r \ll
m^{-1}$. Furthermore, for $r \gg \rho$, $|w_0| \ll 1$ and the field
equations linearize for any $m$. Hence if $m\rho \ll 1$ a smooth field
configuration which is an approximate solution for all $r$ is
\begin{equation}
w_m(x,\tau) = m\rho\, e^{-i\theta} K_1 (mr)\,,
\label{sec:Bes}
\end{equation}
where the Bessel function $K_1$ is the solution of the linearized
Euler-Lagrange equations obeying the boundary condition at infinity.
This approximate solution with action close to the instanton action
$4\pi/g^2$ contributes to the winding number vacuum to vacuum
transition amplitude just as the exact instanton (\ref{sec:inst})
solution does in the symmetric $m=0$ model.
 
In either the softly broken or unbroken sigma model
instanton/anti-instanton configurations are also approximate solutions
to the field equations with finite energy. In the low energy regime
perturbation theory is valid and one can construct the approximate
solutions by forming an infinite chain of alternating instantons and
anti-instantons along the Euclidean time axis separated by a half
period $\beta/2$.  At this point a new parameter, the period $\beta$
enters the problem and we have a finite interaction between instantons
and anti-instantons along the chain, just as in the pendulum example
of the previous section. Hence we can have a competition between the
tendency of the scale $\rho$ to shrink to zero for an instanton in
isolation and the tendency of $\rho$ to increase due to the attractive
interaction between neighboring instantons and anti-instantons, with
the two effects balanced at a particular $\rho$ which is a function of
$\beta$. In order to obtain some detailed understanding of this
qualitative picture and provide a quantitative benchmark for the
numerical methods to follow, we work out in the next section the
periodic instanton solution for energies $E \ll E_{sph}$ where
perturbative methods are applicable.

\section{Low Energy Perturbation Theory}
\label{sec:level4}

Our construction of the periodic instanton solution at low energies
will take place in two steps. Since we are not in possession of an
exact solution to the classical equations for finite $m\rho$ we
evaluate the action of the configuration (\ref{sec:Bes}) to first
non-vanishing order in $m\rho$, at first without regard to the finite
periodicity $\beta$. Then we proceed to patch the configurations
together to construct a trial configuration periodic in Euclidean time
as in the pendulum example. Extremizing the resulting action function
with respect to the free scale parameter $\rho$ will determine the
particular $\rho (\beta)$ at which the attractive interaction between
instantons and anti-instantons just balances the self-interaction
which would cause an instanton to shrink to zero size in isolation.
Provided that in the end $m\rho(\beta)\ll 1$, the expansion is
consistent and we may trust the resulting extremal action constructed
in this way as the correct periodic instanton action to first
non-vanishing order in the small parameter $m\rho(\beta)$, or
equivalently $E/E_{sph}$.

The sigma model action for any function of the form $w(x,\tau) =
e^{-i\theta}f(r)$ may be expressed as 
\begin{equation}
S = {4\pi\over g^2}\int_0^{\infty}r\,dr\, \left[{\left(f'^2 +f^2/
r^2\right)\over\left(1+f^2\right)^2} + {m^2
f^2\over\left(1+f^2\right)}\right]~.
\label{sec:instact0}
\end{equation}
For the trial configuration $f = m\rho K_1(mr)$ we make use of the
Bessel function identity, $K_1' + K_1/u = -K_0$ to rewrite the first
integral in the preceding expression in the form,
\begin{equation}
m^2\rho^2\int_0^{\infty}u\,du\, {\left(K_0^2(u) -2K_1K_1'/u
\right)\over\left[1+ m^2\rho^2K_1^2(u)\right]^2}\,.
\label{sec:firstint}
\end{equation}

Let us divide the integration over $u\equiv mr$ into two parts, $u\le
a$ and $u\ge a$ where $a$ is chosen so that $m\rho \ll a \ll 1$ in the
limit of small $m\rho$. For $u\le a$ we can make use of the series
expansions of the Bessel functions for small argument, 
\begin{eqnarray}
K_0(u) &=& -\left[\ln\left({u\over 2}\right)+ \gamma_E\right]\left(1 +
{\cal O}(u^2)\right)~, \nonumber\\ 
K_1 (u) &=& {1\over u} + {u\over 2}\left[\ln\left({u\over 2}\right)+
\gamma_E - {1\over 2}\right]\left(1 + {\cal O}(u^2)\right)~,\nonumber\\ 
K_2(u) &=& {2\over u^2} - {1\over 2} + {\cal O}(u^2 \ln u)~,
\end{eqnarray}
taking care to include all terms that give a contribution to the final
answer up to order $m^2\rho^2$, while in the interval $u \ge a$ we can
safely replace the denominator in (\ref{sec:firstint}) by unity since
the terms neglected are higher order in $m^2\rho^2$ and the integral
with lower cutoff $a$ is non-singular. In this way we find
\begin{eqnarray}
I_1 &\equiv& \int_0^{a}u\,du\, {\left(K_0^2(u) -2K_1K_1'/u
\right)\over\left[1+ m^2\rho^2K_1^2(u)\right]^2}\nonumber\\ 
&=& -{1\over a^2} + {1\over m^2\rho^2} - \ln \left({a\over 2}\right) -
\gamma_E + {1\over 2} \label{sec:firsteval}\\
& &+ {\cal O}(m^2\rho^2\ln^2 m\rho, a^2\ln^2 a)\nonumber
\end{eqnarray}
and
\begin{eqnarray}
I_2 &\equiv& \int_a^{\infty}u\,du\, {\left(K_0^2(u)
-2K_1K_1'/u\right)\over\left[1+ m^2\rho^2K_1^2(u)\right]^2}\nonumber\\
&=& {1\over a^2} + \ln \left({a\over 2}\right) + \gamma_E
+ {\cal O}(m^2\rho^2\ln^2 m\rho, a^2\ln^2 a)
\label{sec:seceval}
\end{eqnarray}
where $\gamma_E= 0.577\dots$ is Euler's constant. In the latter
integral we have made use of the integration formula $5.54(2)$ of
Ref. \cite{GR},
\begin{equation}
\int_a^{\infty} u\, du K_0^2 (u) = -{a^2\over 2}\left[ K_0^2(a) -
K_1^2(a)\right]\,.  
\end{equation}
Combining the results (\ref{sec:firsteval}) and (\ref{sec:seceval}),
the $a$ dependence drops out (as it must) and we obtain
\begin{equation}
I_1 + I_2 = {1\over m^2\rho^2} + {1\over 2} + {\cal O}(m^2\rho^2\ln^2 
m\rho)\,. 
\end{equation}
Analyzing the second integral in (\ref{sec:instact0}) in the same way
we obtain
\begin{eqnarray}
I_3 &\equiv&\int_0^{a}{u\,du\,K_1^2(u)\over
\left[1+m^2\rho^2K_1^2(u)\right]}\nonumber\\ 
&=& \ln a - \ln m\rho + {\cal O}(m^2\rho^2\ln^2 m\rho, a^2\ln^2 a)
\end{eqnarray}
and
\begin{eqnarray}
I_4 &\equiv&\int_a^{\infty}{u\,du\,K_1^2(u)\over
\left[1+m^2\rho^2K_1^2(u)\right]}\nonumber\\ 
&=& -\ln \left({a\over2}\right) - \gamma_E - {1\over 2} + {\cal
O}(m^2\rho^2\ln^2 m\rho, a^2\ln^2 a) 
\end{eqnarray}
where in this latter integral we have made use of the integration
formula, 
\begin{equation}
\int_a^{\infty} u\, du K_1^2 (u) = {a^2\over 2}\left[ K_0(a) K_2(a) -
K_1^2(a)\right]\,. 
\end{equation}
Hence we find
\begin{equation}
I_3 + I_4 = - \ln \left({m\rho\over 2}\right) - \gamma_E - {1\over 2}
+ {\cal O}(m^2\rho^2\ln^2 m\rho)\,, 
\end{equation}
and all together the action of the trial configuration is 
\begin{eqnarray}
&&S[w_m] = {4\pi m^2\rho^2\over g^2}(I_1 + I_2 + I_3 + I_4)
\label{sec:beind}\\ 
&&= {4\pi\over g^2}\left\{ 1 - m^2\rho^2\left[\ln\left({m\rho\over
2}\right) + \gamma_E \right] + {\cal O}(m^4\rho^4\ln^2
m\rho)\right\}\nonumber\,.                  
\end{eqnarray}
The leading $4\pi/g^2$ term is clearly the one instanton action of the
unbroken ($m=0$) model, so that in considering periodic instantons
next we should multiply the result (\ref{sec:beind}) by two.

In order to calculate the corrections to the action due to the finite
periodicity in the sigma model we follow the same method as in the
pendulum example. We consider the periodic configuration,
\begin{equation}
w(\tau, x) = \left\{ \begin{array}{ll}
w_m(\tau, x)\,, & \qquad 0\le\tau\le \beta/4\\
w_m({\beta\over 2} -\tau, x)\,, & \qquad \beta/ 4\le\tau\le 3\beta/
4\\ 
w_m(\tau - \beta, x)\,, & \qquad  3\beta/ 4\le\tau\le\beta
\end{array}\right. 
\label{sec:tryw}
\end{equation}
and evaluate its action in the fundamental interval $[0,\beta]$
to first non-trivial order in $m^2\rho^2$. Starting from
\begin{equation}
S = {2\over g^2}\int_0^{\beta}d\tau\int_{-\infty}^{\infty} dx
\left\{ {\vert\partial_{\tau} w\vert^2  + 
 \vert\partial_x w\vert^2\over (1 + \vert w\vert^2)^2}
+ m^2 {\vert w\vert^2 \over 1 + \vert w\vert^2}\right\}
\end{equation}
the $\beta$ dependence of the action comes from the tail of the
configuration where $|{w}| \ll 1$, and hence to leading order in
$m^2\rho^2$ the $|{w}|^2$ terms in the denominators may be
neglected. By introducing the spatial Fourier transform
\cite{Bateman},
\begin{eqnarray}
&&\tilde w_m(\tau, k) \equiv \int_{-\infty}^{\infty} dx\, e^{ikx}
w_m(\tau, x) \nonumber\\ &=&-2im\rho \int_{0}^{\infty} {dx\over r}\,
\left(\tau \cos kx -x
\sin kx\right) 
K_1 (mr)\nonumber\\
&=& -i{\pi\rho\over \omega_k} (\omega_k - k)\exp
(-\vert\tau\vert\omega_k) 
\label{sec:wtrans}
\end{eqnarray}
we find that the $\beta$ dependent part of $S$ may be expressed in the
form,
\begin{eqnarray}
{\partial S\over\partial \beta} &=& {8\over g^2}{\partial
\over\partial \beta}\int_0^{\beta/4}d\tau\int_{-\infty}^{\infty}
{dk\over 2\pi} 
\left\{ \big\vert\partial_{\tau}\tilde w_m(\tau, k)\big\vert^2
+ \right.\nonumber\\
&&\qquad\left.\omega_k^2 \big\vert \tilde w_m(\tau,
k)\big\vert^2\right\}\nonumber\\ 
&=& {4\pi\rho^2\over g^2} \int_0^{\infty} dk\ (\omega_k^2 + k^2)\
\exp \left(-{\beta\omega_k\over 2}\right)
\end{eqnarray}
to leading order in $m^2\rho^2$. The infinite periodicity,
$\beta$ {\em in}dependent part of $S$ cannot be evaluated from only
the quadratic terms in the action but it is just given by twice the
zero energy action we calculated in (\ref{sec:beind}) above. Hence, 
\begin{eqnarray}
&&S[w] = {8\pi\over g^2}\left\{ 1 -
m^2\rho^2\left[\ln\left({m\rho\over 2}\right)+ \gamma_E \right]
\right.\nonumber\\ 
&&\left.- \rho^2\int_0^{\infty} dk {(\omega_k^2 + k^2)\over
\omega_k} \exp\left(-{\beta\omega_k\over 2}\right)
+ {\cal O}(m^4\rho^4\ln^2 m\rho)\right\}\nonumber\\
\label{sec:lowestS}
\end{eqnarray}
is the action of the configuration (\ref{sec:tryw}) to this order.

As in the pendulum example of the last section the joining of the
configurations at $\tau = \beta/4$ and $\tau = 3\beta/4$ produces
delta function source terms proportional to
\begin{eqnarray}
&&L[w] \equiv  \left( - {\partial^2\over \partial\tau^2} -
 {\partial^2\over \partial x^2} 
 + m^2\right) w(\tau, x) = \nonumber\\
&& -2 \partial_{\tau} w_m ({\beta\over 4}, x)
\delta \left(\tau - {\beta\over 4}\right) + 2 \partial_{\tau} w_m
 (-{\beta\over 4}, x) 
\delta \left(\tau - {3\beta\over 4}\right)\,.\nonumber\\
\label{sec:linw}
\end{eqnarray}
Making use of the periodic thermal propagator in two dimensions,
\begin{eqnarray}   
&&G_{\beta}(\tau, \tau'; x, x') = \nonumber\\
&&\int_{-\infty}^{\infty} {dk\over 2\pi}\,{1\over
2\omega_k}{e^{ik(x-x')}\over \sinh \left({\beta\omega_k /
2}\right)} \cosh \omega_k\left(\vert \tau-\tau'\vert -
{\beta\over2}\right) 
\label{sec:therGo}
\end{eqnarray}
and taking account of the two real degrees of freedom in the complex
field $w$, we obtain the quadratic shift in the action,
\begin{equation}
{2\over g^2}\int d\tau\int d\tau' 
\int dx\int dx'L'[w^*]G_{\beta}(\tau, \tau'; x, x') L[w]\,.
\label{sec:shiftact}
\end{equation}
Substituting  (\ref{sec:linw}) and (\ref{sec:therGo}) into this
expression, we find that the quadratic shift becomes
\begin{eqnarray}
&&{4\over g^2}\int_0^{\infty} {dk\over \pi\omega_k}\left\{\big\vert 
\partial_{\tau}\tilde w_m \left({\beta\over 4},
k\right)\big\vert^2\,{\rm coth}  
\left({\beta\omega_k\over 2}\right)\right.\nonumber\\
&&\left.- {1\over \sinh
\left({\beta\omega_k/2}\right)}\partial_{\tau}\tilde
w_m^*\left({\beta\over 4}, k\right)  
\partial_{\tau}\tilde w_m\left(-{\beta\over 4}, k\right) \right\}\,.
\end{eqnarray}
We did not need to keep the analog of the second term above in the
pendulum model since there it is subdominant to the first term as $\beta
\rightarrow \infty$.  However, in the $O(3)$ model $\beta \rightarrow
0$ as $E \rightarrow 0$ as we shall see presently, and this second
term must be retained as well.

Since from (\ref{sec:wtrans}),
\begin{equation}
\partial_{\tau}\tilde w_m\,\left(\pm{\beta\over 4}, k\right) = i\pi
\rho\,(\omega_k \mp k)\, \exp \left(-{\beta\omega_k\over 4}\right)\, ,  
\end{equation} 
the quadratic correction to the action arising from the shift
(\ref{sec:shiftact}) is
\begin{eqnarray}
&&- {8\pi\rho^2\over g^2}\int_0^{\infty}{dk\over\omega_k}{1\over \sinh
\left({\beta\omega_k / 2}\right)}\times\nonumber\\
&&\left[k^2 \left( 1 + e^{-\beta\omega_k}\right) 
- {m^2 \over 2}  \left( 1 - e^{-\beta\omega_k}\right)\right]\,.
\end{eqnarray}
Combining this shift with the unshifted action (\ref{sec:lowestS}) we
secure finally,
\begin{equation}
S(\beta, \rho) = {8\pi\over g^2}\left\{1 - m^2\rho^2\left[F(m\beta) +
\ln\left({m\rho\over 2}\right)+ \gamma_E \right]\right\} \label{rhoind}
\end{equation}
to the lowest non-trivial order in $m\rho$, where 
\begin{equation}
F(m\beta) \equiv \int_0^{\infty}{dk\over\omega_k}{1\over \sinh
\left({\beta\omega_k / 2}\right)}\left[{2 k^2\over m^2} + 1 -
e^{-{\beta\omega_k\over 2}}\right] > 0\,. 
\end{equation}
This result may also be obtained by the ``R-term'' method
\cite{KRT,RST}. 

With the action of the trial configuration in hand for arbitrary small
$m\rho$ we can determine the value of $\rho$ which extremizes the
action and therefore leads to a classical periodic instanton solution
close to the trial configuration (\ref{sec:tryw}). Since
\begin{equation}
{\partial S(\beta, \rho)\over \partial \rho} = 
-{8\pi m^2\over g^2}\rho \left[ 2F + 2\ln\left({m\rho\over 2}\right) +
2\gamma_E + 1\right] 
\label{sec:rhocr}
\end{equation}
vanishes for $\rho =0$ or for
\begin{equation}
\rho = \rho (\beta) = {2\over m} \exp \left(- F(m\beta) -\gamma_E -
{1\over 2} \right)~, 
\label{sec:lowErho}
\end{equation}
the zero energy singular instanton at $\rho = 0$ with finite action $S
= 8\pi/g^2$ is always a solution of the equations for any period
$\beta$, but there is also a non-trivial periodic instanton solution
with action,
\begin{equation}
S(\beta) = S(\beta, \rho(\beta)) = {8\pi\over g^2}\left[1 + 2
 e^{-2F(m\beta)- 2\gamma_E - 1}\right] > {8\pi \over g^2}
\label{sec:Spinst}
\end{equation}
and energy,
\begin{eqnarray}
E(\beta) &=& {dS \over d\beta}\nonumber\\
&=& - {32\pi m\over g^2}
e^{-2F(m\beta)- 2\gamma_E - 1} F'(m\beta) \nonumber\\   
&=& {16\pi e^{-2\gamma_E -1} \over g^2 }e^{-2F}\int_0^{\infty}\,
{dk\over \sinh^2 ({\omega_k\beta / 2})}\times\nonumber\\
&&\left\{ \left({2k^2\over m^2} +
1\right)\cosh\left({\omega_k\beta \over 2}\right) - 1 \right\}\,.   
\label{sec:lowE}
\end{eqnarray}
This second non-singular solution with non-zero $\rho (\beta)$ is the
periodic instanton solution we seek in the limit of small but finite
$E$.  

The qualitative picture sketched in the previous section has been
verified by this explicit calculation, namely at the non-trivial value
of $\rho (\beta)$ in (\ref{sec:lowErho}) the tendency of an individual
instanton or anti-instanton to contract to $\rho =0$, represented by
the logarithm in (\ref{sec:rhocr}) is just balanced by the attractive
interaction between members of the infinite chain which tends to
increase $\rho$, represented by the $F(m\beta)$ term in
(\ref{sec:rhocr}). This competition between the two effects can occur
in the sigma model only because of the softly broken conformal
invariance of the mass term $S_m$ which leads to the existence of the
conformal scale parameter $\rho$.

Notice also that in contrast to the pendulum example of the last
section it is {\em not} $\beta \rightarrow \infty$ which characterizes
the low energy limit $E \rightarrow 0$, but rather $F(m\beta)
\rightarrow \infty$ in the sigma model. Since $F(m\beta)$ vanishes
exponentially as $\beta \rightarrow \infty$ but in the opposite limit,
\begin{equation}
F(m \beta)\rightarrow {\pi^2 \over m^2\beta^2}\rightarrow \infty
\qquad {\rm as} \qquad m\beta \rightarrow 0\,, 
\end{equation}
the low energy limit is characterized by $\beta \rightarrow 0$.  This
(possibly counterintuitive) result is nevertheless consistent with
perturbation theory and the dilute gas picture of instantons and
anti-instantons at low energies because the size of the individual
(anti-)instantons goes to zero much more rapidly with $\beta
\rightarrow 0$, and it is the ratio,
\begin{equation}
{\rho (\beta)\over \beta} \rightarrow {2e^{-\gamma_E - {1\over
2}}\over m\beta} \exp \left(- {\pi^2 \over m^2\beta^2}\right)
\rightarrow 0  
\end{equation}
which controls the validity of the dilute gas approximation.  Clearly
then as $\beta \rightarrow 0$, $\rho(\beta) \rightarrow 0$
exponentially, and it is legitimate to neglect the higher powers of
$m\rho$ in (\ref{sec:lowE}) in the low energy limit, $E \ll E_{sph}$,
which justifies our expansion in powers of $m \rho$ {\em a
posteriori}.  In the extreme low energy limit we have the asymptotic
forms,
\begin{eqnarray}
S(\beta) &\rightarrow& {8\pi\over g^2}\left[1 + {2\over e^{2\gamma_E
+1}} \exp\left(-{2\pi^2 \over m^2\beta^2}\right)\right]\,,\qquad{\rm
and}\nonumber\\ 
E(\beta) &\rightarrow& {64\pi^3\over g^2e^{2\gamma_E +1}m^2\beta^3} 
\exp\left(-{2\pi^2 \over m^2\beta^2}\right)\,.
\end{eqnarray}

{}From either this last expression or the more accurate
(\ref{sec:lowE}) above we find that the behavior of $E$ increasing
with increasing $\beta$,
\begin{equation}
{d E \over d\beta} > 0
\end{equation}
is actually the case for the periodic instantons in the low energy
limit of the nonlinear sigma model. We shall see in the next section
that this behavior persists for all energies up to the sphaleron
energy. Hence it is already clear that the behavior of the periodic
instantons in the $O(3)$ model is very different from that of systems
with only one quantum mechanical degree of freedom such as the simple
pendulum, or the abelian Higgs model. In those models $\beta(E)
\rightarrow \infty$ as $E\rightarrow 0$, and the periodic instanton
contributes to the transition rate as the temperature $\beta^{-1}$
goes smoothly to zero. From the discussion at the end of the second
section it follows that these periodic instantons in the sigma model
with period going to {\em zero} as $E\rightarrow 0$ cannot contribute
to the low temperature transition rate, although they can contribute
to, and in fact dominate, transitions between {\em non}-thermal low
{\em energy} states.

Clearly this behavior is a direct consequence of the conformal
invariance of the unbroken model which leads to the additional scale
parameter $\rho$ in the instanton configuration, and as it turns out,
an additional negative mode direction in the fluctuations around this
configuration. This additional negative mode is easily checked from
the second derivative of the classical action (\ref{rhoind}),
\begin{eqnarray}
&&{\partial^2 S \over \partial\rho^2}\Big\vert_{\rho =\rho(\beta)} =
\nonumber\\
&&-{8\pi m^2\over g^2}\left(3 + 2\gamma_E + 2 \ln \left({m\rho \over
2}\right)+ 2F(m\beta)\right)\Big\vert_{\rho =\rho(\beta)}\nonumber\\ 
&& = -{16\pi\over g^2}m^2 < 0
\label{sec:negrho}
\end{eqnarray}
on the solution with non-zero $\rho (\beta)$. Thus, any small change
of $\rho$ from its critical value, $\rho(\beta)$ {\em decreases} the
action of the configuration and implies the existence of a negative
eigenvalue in the second order fluctuation operator around the
non-singular periodic instanton solution that is quite independent of
the usual negative mode with fixed $\rho$ as found by the standard
argument of the Section 2. The origin of this second negative
mode about the non-trivial periodic instanton solution we have found
for low $E$ should be clear from what has already been said regarding
the balance between the attractive interaction between nearest
neighbor instantons and anti-instantons on the one hand, and the
self-interaction attracting each individual (anti-)instanton towards
zero size on the other. If the precise balance between the two is
upset by varying $\rho$ slightly away from its critical value,
$\rho(\beta)$, then one or the other of the two interactions dominates
and $\rho$ is driven further away from its critical value, {\em i.e.}
there is a negative mode in the direction of varying $\rho$, which is
just what (\ref{sec:negrho}) makes explicit.

%FIG5 (INVMH)
\vspace{.4cm} 
\epsfxsize=6cm
\epsfysize=4cm
\centerline{\epsfbox{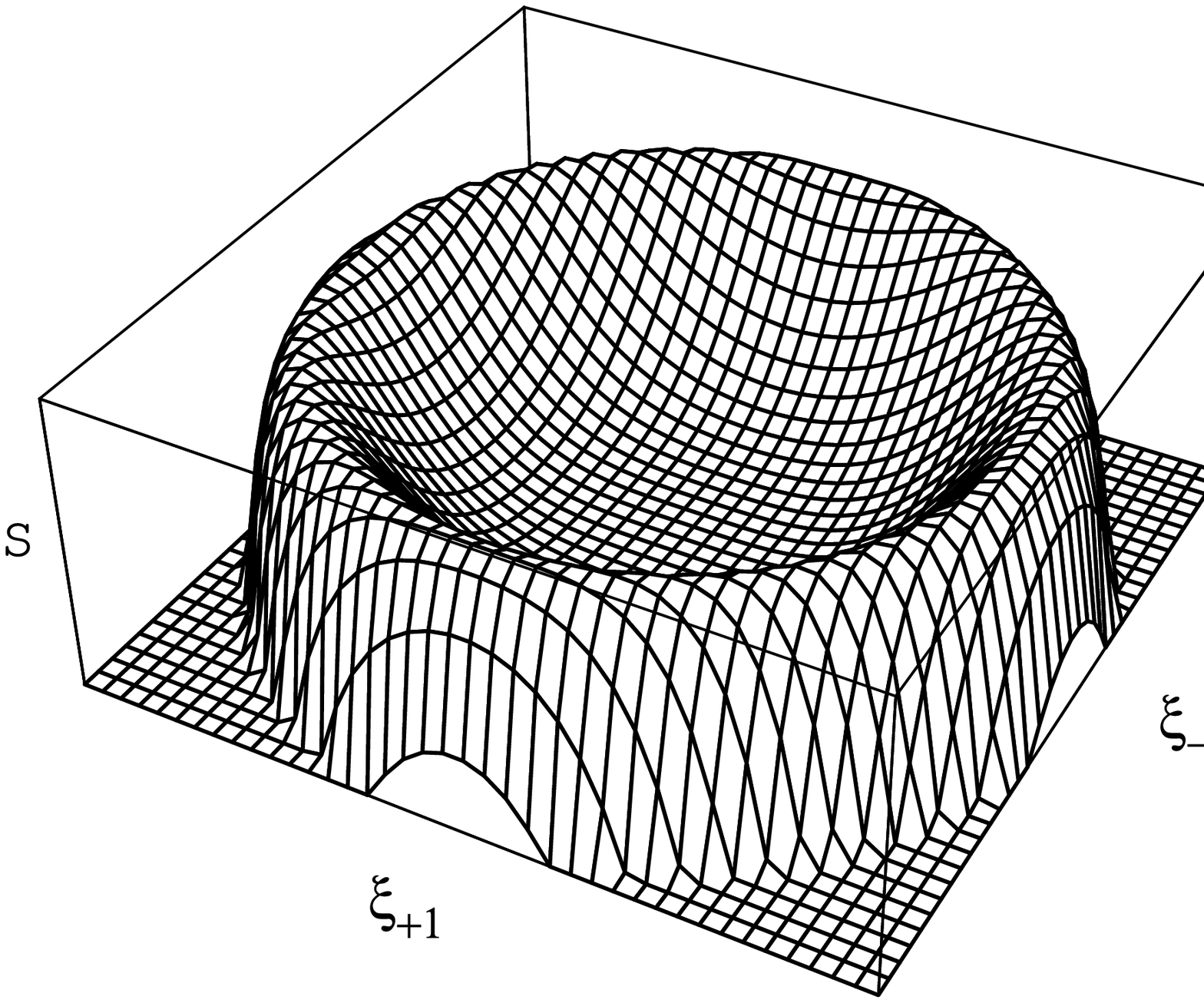}}
\vspace{.35cm}
\epsfxsize=6cm
\epsfysize=4cm
\centerline{\epsfbox{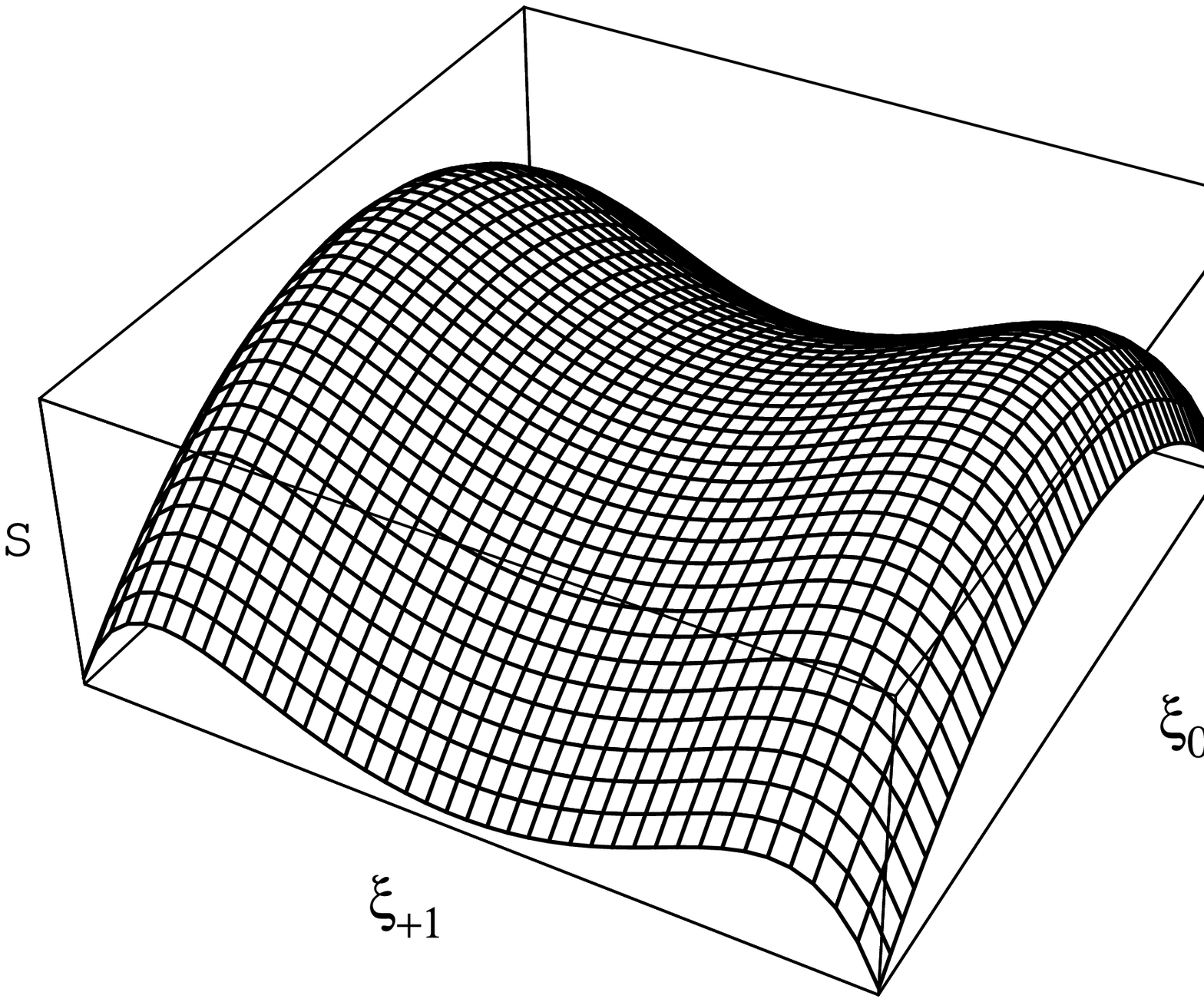}}
\vspace{.35cm}
{FIGS. 5. {\small{The action $S$ plotted as a function of the
same variables $\xi_{\pm 1}$ (a), and $\xi_0$ (b), for fixed $\beta < \beta_-$,
as in Figs. 3, but for the $O(3)$ sigma model. The periodic instanton
is again represented by the ring of extrema of $S$ in (a) or
the two outer extrema in (b), which merge with the sphaleron in the
center as $\beta$ approaches $\beta_-$. Since the periodic instanton
solution now has two negative modes and higher action than the sphaleron,
it is subdominant. The finite temperature winding number transition goes
over the lowest saddle, either the sphaleron or the $\rho=0$ instanton
(not shown) depending on the temperature.}}}\\ 

It is this second negative mode direction due to conformal rescalings
that makes the non-trivial periodic instanton configuration
subdominant at finite temperature, at least in the low energy limit
where perturbation theory and the form (\ref{sec:lowE})
applies. Correspondingly, the periodic instanton action
(\ref{sec:Spinst}) is {\em greater} than the singular solution with
$\rho = 0$. Near the sphaleron energy the action as a function of the 
three relevant variables $\xi_{\pm 1}$ and $\xi_0$ for the sigma model 
is illustrated in Figs. 5. 

{}From the first form of Eqn. (\ref{sec:negrho}) we observe that
the zero size instanton/anti-instanton configuration has no second
negative mode in the $\rho$ direction. It has only the expected single
negative mode corresponding to the attractive interaction between the
pair. Hence this singular configuration with fixed finite action per
period $2S_0$ (shown as the constant straight line in Fig. 4) can and
does contribute to the winding number transition rate at low
temperature $T<T_{cr}$. Notice from Fig. 4 that although the actions
$2S_0$ and $E_{sph}\beta$ are equal at $\beta=1/T_{cr}$ the actual
field configurations of the zero size instanton/anti-instanton pair
and the sphaleron are quite different and there is no merging of these
solutions at $\beta=\beta_{cr}$.
 
The picture we have sketched in this section is based on the existence
of a perturbative instanton/anti-instanton expansion at low energies
and knowledge of the sphaleron and its negative mode at finite energy
and $\beta$ near $\beta_{cr}$. Continuity of the space of solutions to
the classical Euclidean equations suggests that the periodic instanton
solutions merge with the sphaleron at $\beta$ approaches $\beta_{cr}$
from below with $dE/d\beta > 0$. However one does not know how the
action and energy of these solutions will vary as $\beta$ is varied
from $\beta_{cr}$ {\em a priori}. In the absence of analytic methods
for finding the periodic instanton solutions to the Euclidean
equations in the intermediate region between $E\rightarrow 0$ where
perturbation theory holds and $E\rightarrow E_{sph}$ where the
solutions approach the static sphaleron, one must rely on a numerical
approach. It is to the details and results of this numerical study
which justifies Figs. 4 and 5 that we turn next.

\section{Numerical Technique and Results}
\label{sec:level5}

In this section we present the results of a numerical study of
periodic instanton solutions in the two dimensional nonlinear $O(3)$
sigma model modified by the mass term $S_m$. A preliminary version of
these results has been reported earlier in Ref. \cite{HMT}.  At
energies comparable, but not very close to the sphaleron energy, the
periodic instanton solution has to be found numerically, {\em i.e.}
one must solve the Euclidean field equations,
\begin{equation}
-(\partial_\tau^2 + \partial_x^2) n^a + m^2 \delta^a_3-\alpha n^a=0,
\label{sec:eqsom}
\end{equation}
together with the constraint,
\begin{equation}
n^an^a=1\,.            
\label{sec:cont eqs}
\end{equation}
Here, $\alpha$ is a Lagrange multiplier enforcing the constraint,
which is easily found by multiplying (\ref{sec:eqsom}) by $n^a$ and
using (\ref{sec:cont eqs}), namely
\begin{eqnarray}
\alpha &=& -n^a (\partial_\tau^2 + \partial_x^2) n^a + m^2
n_3\nonumber\\ 
&=& \partial_{\mu}n^a\partial_{\mu}n^a + m^2 n_3\,.
\end{eqnarray}
The periodic instanton solution we seek has vanishing time derivative
at initial time $\tau=0$, evolves to another turning point at
half-period $\beta/2$ where it reflects and then returns to its
initial configuration at $\tau = \beta$ by simply reversing the sign
of all $\tau$ derivatives. Thus we enforce the half-period boundary
conditions,
\begin{equation}
{\partial n^a(\tau=0,x)\over\partial \tau}={\partial
n^a(\tau=\beta/2,x)\over\partial \tau}=0 \ , 
\end{equation}
which removes the time translation invariance of the solution. The
other boundary condition we require is that at spatial infinity the
solution approaches the vacuum. Since we shall work in a finite box of
length $L$ we require
\begin{equation}
n^a(\tau, x=-L/2)= n^a(\tau, x=L/2) = n_{(vac)}^a = (0,0,-1)
\end{equation}
for $mL$ large but finite.

%PUT THE RECTANGLE HERE (FIG6)
\vspace{.4cm} 
\epsfxsize=6cm
\epsfysize=4cm
\centerline{\epsfbox{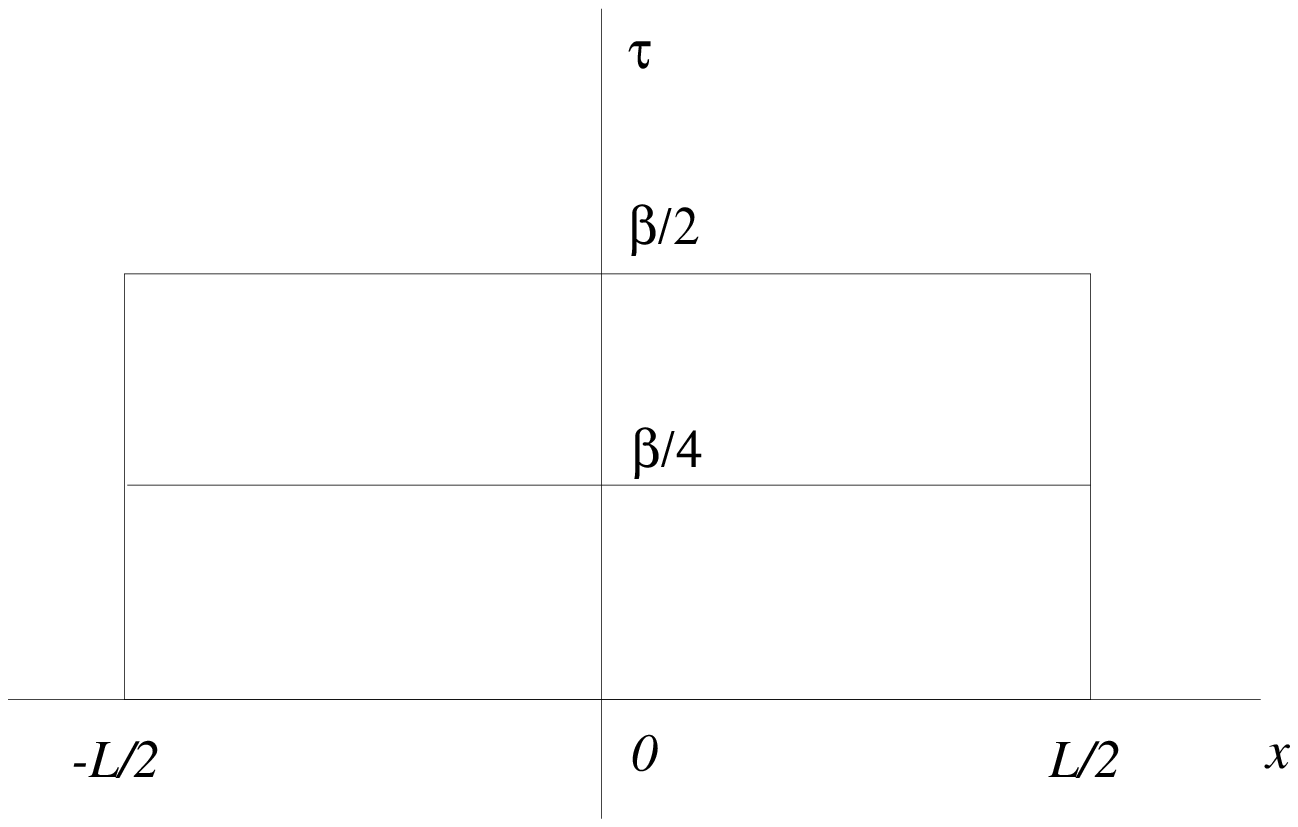}}
\vspace{.35cm}
{FIG. 6. {\small{The finite rectangular region $(\tau,x)$ of
coordinate space that maps onto the shaded region of the sphere in
Fig. 7.}}}\\

%PUT THE BALL HERE (FIG7)
\vspace{.35cm} 
\epsfxsize=6cm
\epsfysize=6cm
\centerline{\epsfbox{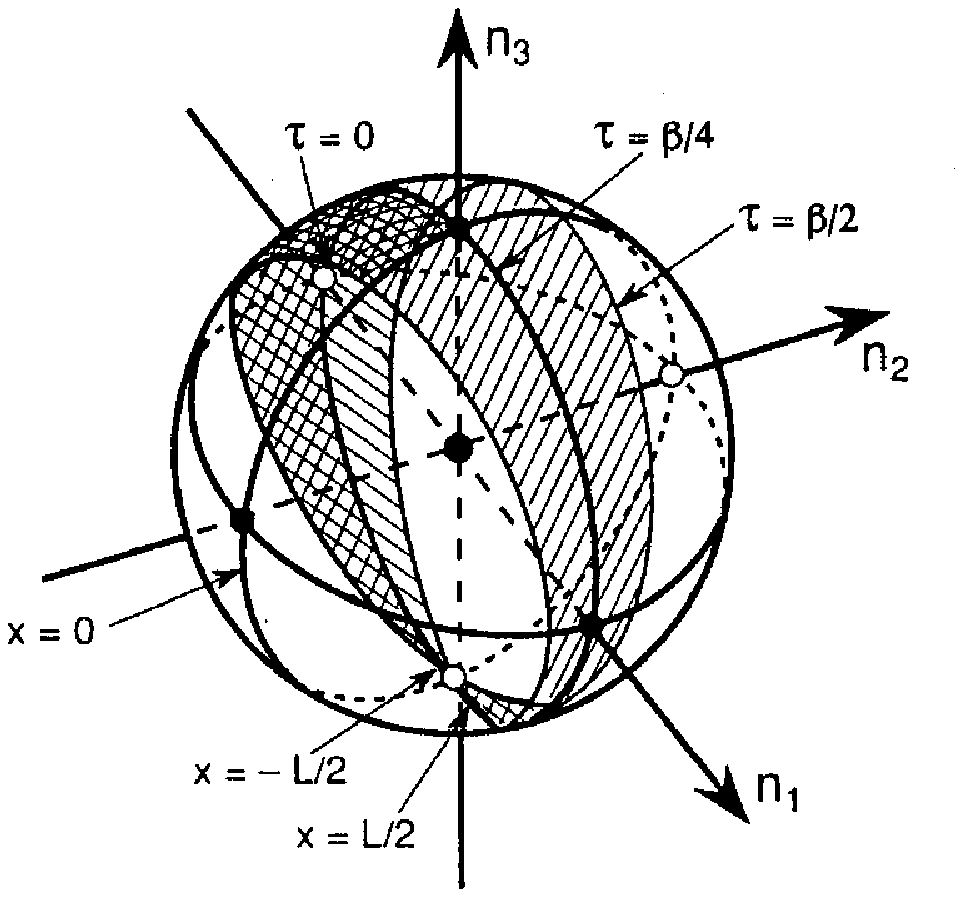}}
\vspace{.1cm}
{FIG. 7. {\small{Geometric representation of the periodic
instanton range on $S^2$.}}}\\

Geometrically, the periodic instanton solution obeying these boundary
conditions maps the rectangular region of $(\tau,x)$ coordinate space
pictured in Fig. 6 into the shaded region of the sphere in
Fig. 7. From these figures it should be clear that the solution may be
chosen to have well-defined symmetry properties under reflection
through the lines bisecting the rectangular region of Fig. 6, so that
at $x=0$, $n_1$ vanishes while $n_2$ and $n_3$ reach extrema, and at
$\tau = \beta/4$, $n_2$ vanishes while $n_1$ and $n_3$ reach
extrema. Imposing these symmetries fixes completely the spatial
translational invariance and rotation invariance around the $n_3$ axis
of the solution, as well as reduces the region we must consider to
only one quarter of the full rectangle in Fig. 6. Hence we solve the
differential Eqns. (\ref{sec:eqsom}) subject to the boundary
conditions,
\begin{equation}
\begin{array}{lcl}
x=-L/2 & : & n_1=0;\;\;n_2=0;\;\;n_3=-1; \\
\tau=0    & : & dn_1/d\tau=dn_2/d\tau=dn_3/d\tau=0;\\
x=0    & : & n_1=0;\;\;dn_2/dx=dn_3/dx=0;\\
\tau=\beta/4  & : & n_2=0;\;\;dn_1/d\tau=dn_3/d\tau=0.
\end{array}                                 \label{sec:bc}
\end{equation}

The lattice version of Eqns. (\ref{sec:eqsom}) can be obtained by a
variational principle starting from the discretized action. Let
$(\tau_i,x_j)$, where $i=0,\ldots, I$ and $j=0,\ldots, J$, be the
coordinates of lattice sites. The lattice version of the action
(\ref{sec:action0}) and (\ref{sec:action1}) reads
\begin{eqnarray}
S&=&\sum\limits_{i,j} \Bigl\{{1\over 2} (n^a_{i+1,j}-n^a_{ij})^2 
{d\tilde{x}_j\over d\tau_i}
+{1\over 2} (n^a_{i,j+1}-n^a_{ij})^2  {d\tilde{\tau}_i\over
dx_j}\nonumber\\ 
&&+ m^2(1+n^3_{ij})d\tilde{\tau}_id\tilde{x}_j \Bigr\}
\label{sec:action discr}
\end{eqnarray}
where we have used the notations $d\tau_i=\tau_{i+1}-\tau_i$,
$d\tilde{\tau}_i=(1/2)(d\tau_{i-1}+d\tau_i)$ and similarly for $dx_j$,
$d\tilde{x}_j$.  

The equations which follow from the action (\ref{sec:action discr}) 
can be written in the form 
\begin{equation}
[B^a-n^a (B^b n_b)]_{ij}= 0, \label{sec:discr eqs}
\end{equation}
where 
\begin{eqnarray}
B^a_{ij}&\equiv& - {n^a_{i+1,j}\over d\tilde{\tau}_id\tau_i}
	- {n^a_{i-1,j}\over d\tilde{\tau}_id\tau_{i-1}}
- {n^a_{i,j+1}\over d\tilde{x}_jdx_j}
- {n^a_{i,j-1}\over d\tilde{x}_jdx_{j-1}}\nonumber\\
&&+ m^2\delta_{a3}. \label{sec:Bij}
\end{eqnarray}
Eqn. (\ref{sec:discr eqs}) actually contains only two independent
equations since its projection onto the vector $n^a_{ij}$ is
identically zero. Together with the constraint equation (\ref{sec:cont
eqs}) these comprise a complete set of three independent equations
associated with each lattice site. From Eqns.  (\ref{sec:discr eqs}) and
(\ref{sec:Bij}) it is clear that $B^a$ must be anti-parallel to $n^a$ 
at each point $(i,j)$. Since $n^a$ is normalized to unity, the 
Eqns. (\ref{sec:discr eqs})--(\ref{sec:Bij}) can be rewritten in the 
equivalent, symmetric form,
\begin{equation}
n^a\sqrt {B^cB^c} + B^a=0,         \label{sec:symm eqs}
\end{equation}
for every $(i,j)$, and where the negative sign of the square root has
to be taken because $n_aB^a < 0$. This form is most convenient for 
numerical calculations.

To obtain a numerical solution of (\ref{sec:symm eqs}) we use Newton's
method \cite{NM}. That is, we \begin{enumerate}
\item choose an initial field configuration as a first guess;
\item linearize the equations (\ref{sec:symm eqs}) in the background of 
the initial field;
\item solve the linearized equations to obtain an improved
configuration; and 
\item iterate until the procedure converges.
\end{enumerate} 
The third step above also involves a renormalization: the improved 
configuration after every Newton step is locally scaled to ensure
that at every lattice point the constraint (\ref{sec:cont eqs})
is satisfied. This procedure speeds up the convergence to the final 
solution. The numerical solutions were found for $m=1=g$.

The choice of initial configuration is guided by the known behavior of
the periodic instanton solution near $E=0$ and $E=E_{sph}$.
Consideration of the zero mode in the vicinity of the sphaleron when
$\beta$ approaches $\beta_-$ indicates that we should look for
non-trivial periodic solutions of the field equations in the
neighborhood of the sphaleron by perturbing the known static solution
along the zero mode direction, {\em viz.}
\begin{equation}
n_{(sph)}^a(x) \rightarrow n_{(sph)}^a(x) + \delta\;
{\cos}(|\omega_-|\tau) u^a_-(x) 
\label{sec:nearEsph} 
\end{equation}
where $\delta$ is a small parameter which goes to zero at $\beta
=\beta_-$.  Since the classical action function on this solution is
identical to Hamilton's principal function, the energy of the solution
is given by
\begin{equation}
{dS(\beta)\over d\beta}=E\,,  \label{sec:energy cond} 
\end{equation}
which must agree with the sphaleron energy $E_{sph}$ at
$\beta=\beta_-$. For $E$ just below $E_{sph}$ an initial configuration
of the form (\ref{sec:nearEsph}) with $\delta\sim 0.5$ works well. The
convergence of the Newton scheme is quadratic, and depending on the
initial starting point, the desired accuracy (an error tolerance of
$10^{-10}$ or better) is typically reached in several iterations. Error 
tolerance limits were set to limit both the maximum violation of the 
equations of motion at a given lattice site as well as the global error 
obtained by summing the absolute value of, and then averaging, the 
errors on all lattice sites. Once the solution has been found at a given 
$\beta$, we step down (or up) in $\beta$ and use the previous solution 
as the initial trial configuration for Newton's method at the next value 
of $\beta$. If the step size in $\beta$ is not too large this procedure
enables us to efficiently generate periodic instanton solutions
over a finite range of $\beta$.

The solution of the linearized equations in step (3) of the algorithm
can be performed by a direct inversion of the matrix of small
fluctuations about the trial configuration. (This was the method used
in our earlier work on this problem \cite{HMT} and is described
briefly below.) Negative eigenvalues of this matrix are treated on the
same footing as positive eigenvalues, and pose no special problem for
Newton's method, which is important for the present
application. Because of the boundary conditions which fix the
translational and rotational symmetries there are no zero eigenvalues
of the matrix near the desired solution, which otherwise would be
disastrous for this method.

For an $I\times J$ spacetime grid the matrix to be inverted has
$I^2J^2$ elements. In general, it takes of order $I^3J^3$ operations
to invert such a matrix. However, the sparseness of the matrix makes
it more efficient to follow a procedure of forward elimination and
back substitution along either the $x$ or $\tau$ directions
instead. That is, starting with one edge of the region in Fig. 6, such
as $\tau =0$, we can solve for each $\tau$ slice of the grid in terms of
the successive two $\tau$ slices until the edge $\tau = \beta/4$ is
reached, where the boundary condition determines the unknown
quantities. Then we reverse direction and solve for the unknowns on
the previous $\tau$ slices successively. This allows for the matrix to
be inverted in order $IJ^3$ operations (or $I^3J$ operations if the
$x$ direction is chosen), and speeds up the algorithm considerably.
In practice, a grid size of order $100\times 100$ can be handled on a
typical workstation, which provides reasonable accuracy for
configurations not too far from the sphaleron. However, this method
rapidly becomes inefficient when large grids are required.

The strategy adopted in this paper is to solve the linearized
equations using direct matrix methods such as conjugate gradient. The
present code allows for the use of several such solvers (all capable
of dealing with non-symmetric, complex matrices). The results reported
here were obtained using either the quasi-minimized version of
Sonneveld's conjugate gradient squared (CGS) algorithm \cite{sonn} or
Van der Vorst's bi-conjugate gradient with stabilization (BICG\_STAB)
\cite{vdv}. The advantage of these methods over matrix inversion is a
much less stringent memory requirement (down by a factor of $I$ (or
$J$) compared to matrix inversion) and while of the same order
algorithmically, typically the prefactor is much smaller. Moreover,
the efficient use of a parallel supercomputer becomes possible.

Global grid refinement was used to speed up the method even further,
as well as to effectively increase the order of the discretization
error. The technique used was to begin with a certain grid size, find
the periodic instanton, and then to double the grid size (keeping the
physical volume constant) and use an interpolated version of the
solution on the smaller grid as the initial guess solution for the
bigger lattice. This procedure works as a sort of preconditioner for
the matrix solver and since one has the results from the coarser
lattice, Richardson extrapolation \cite{richext} can be used to reduce
the discretization error, effectively improving the second order
finite differencing error to third order (Fig. 8). A good check of the
error control achieved by the code is energy conservation. Energy
error is maximum near $\tau\sim\beta/4$ where the solution is far from
the vacuum and varying rapidly in both time and space. An example of
the improvement in energy error as a function of grid size (before
Richardson extrapolation) is shown in Fig. 9.

In practice, the number of (matrix solver) iterations hardly ever went
beyond tens of thousands even for grids as big as of order $1000\times
1000$ and acceptable accuracies (error tolerances of order $10^{-10}$)
were reached in less than four doubling steps starting with
configurations of order $100\times 100$.  When doing runs for
different values of $\beta$, the solution for the previous $\beta$ on
the biggest grid was sampled to produce a guess solution for the new
$\beta$ on the smallest grid. This allowed efficient scanning of the
desired range of $\beta$ values. As will be seen below the numerical
results are extremely accurate all the way down to small values of
$\beta$ where the perturbative results of the previous section begin
to apply.

%INSERT FIG8 HERE (NEWACTRICH)
\vspace{.4cm} 
\epsfxsize=6cm
\epsfysize=5cm
\centerline{\epsfbox{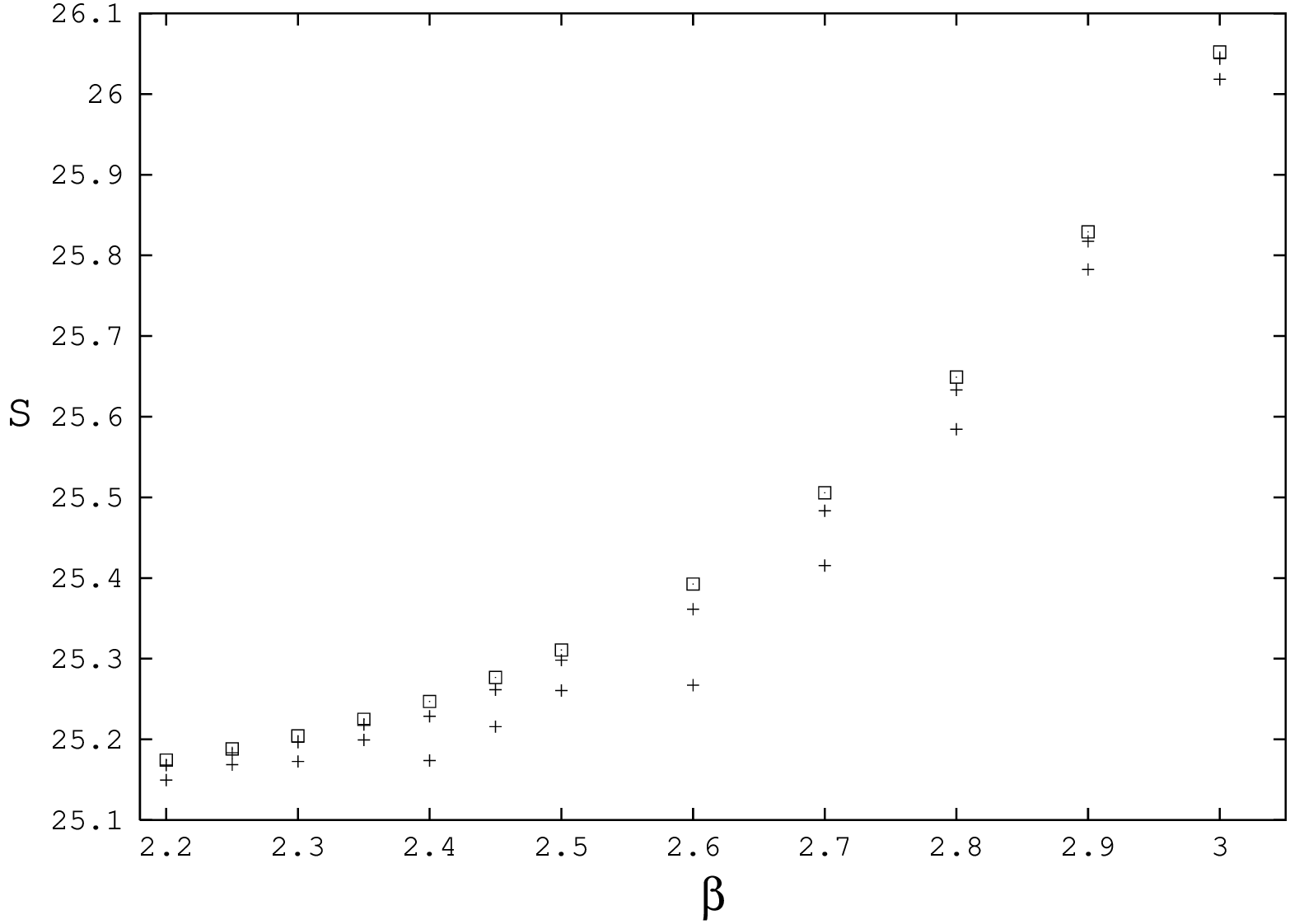}} 
\vspace{.35cm}
{FIG. 8. {\small{Action of the periodic instanton as
a function of $\beta$ illustrating the improvement due to
extrapolation. The two crosses at each $\beta$ are the numerical
results from the largest and next-to-largest lattice at that $\beta$
(top and bottom, respectively). The square is the extrapolated
point. Accuracy of the extrapolation was checked by extrapolating up
from a still coarser, {\em i.e.} next-to-next-to-largest lattice and
comparing with the actual result at the largest lattice: there was
very good agreement in all cases.}}}\\ 

%INSERT FIG9 HERE (NEWENCHECK)
\vspace{.35cm} 
\epsfxsize=6cm
\epsfysize=5cm
\centerline{\epsfbox{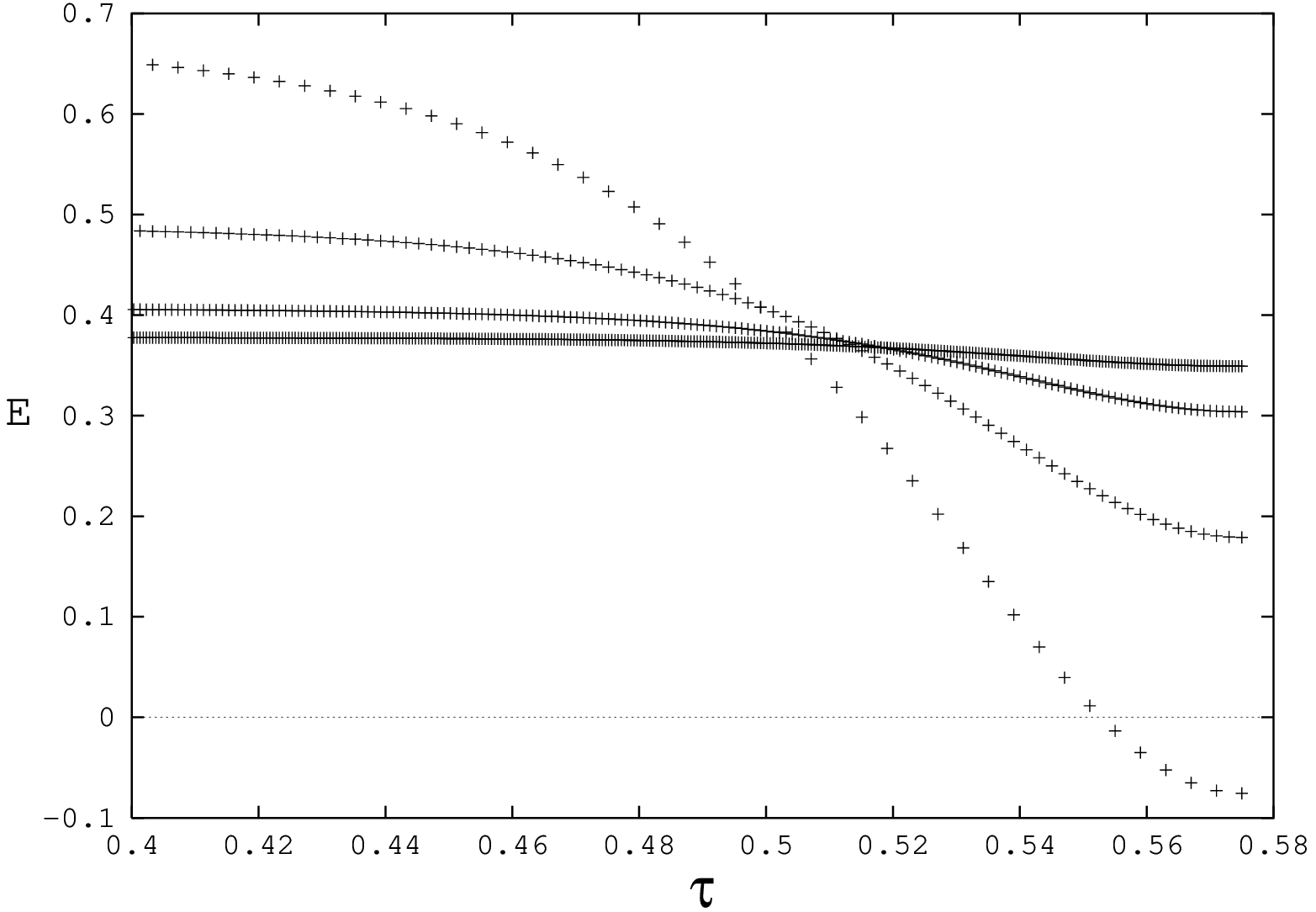}} 
\vspace{.35cm}
{FIG. 9. {\small{Energy of the periodic instanton as
a function of the time-slice at $\beta=2.3$. A small range of $\tau$
is shown to illustrate the improvement in energy conservation as the
grid is refined (over four doublings of the grid in this case).
Extrapolation improves the result one step further (not shown).}}}\\

Without any further optimizations, the new method is already efficient
and accurate, yielding a real periodic instanton solution for energies 
in the range of $.3$ to $8$ in units of $m/g^2$. Smaller energies 
require finer lattices to obtain the same accuracy, since the instanton 
size becomes smaller rapidly with decreasing $\beta$. Adaptive gridding
using either nonuniform grids or selective refinement is possible
within the method but was not used as a meaningful comparison with
perturbation theory was already possible with the largest grid sizes
that were used. (A selective refinement algorithm has been implemented
for application to future, more demanding problems.)

The results of the calculations are summarized below. They were
obtained on grids as large as $\sim 1200\times 2400$ in spatial boxes
with sizes varying from $L=8$ to $L=16$ in units of $m^{-1}$ (for
configurations nearer to the sphaleron smaller grids provided
sufficient accuracy). For $\beta > \beta_-$, as discussed in the
previous section, a real solution does not exist and the solution
becomes complex. This complex solution was also found without
difficulty by our numerical method. Numerically, the crossover from
the real to complex solution occurred between $\beta=3.62$ (real) and
$\beta = 3.63$ (complex) consistent with $\beta_-=3.628$ (Fig. 4). The
action of the complex solution rises steeply at $\beta\approx 4$ and
it becomes difficult to track this numerical solution for larger
$\beta$ (possible in principle, but not worth the computer time). The
numerical results for action and energy versus $\beta$, and $W$ as a
function of $E$ are given in Figs. 10. On the scale of the plots, the
numerical error estimates are insignificant and are not shown.
Comparison with the perturbative results of the previous section shows
very good agreement at low values of $\beta$ as expected. This is also
illustrated in Figs. 10 by plotting $S$, $E$, and $W$ over a
restricted range within which perturbation theory is supposed to
become accurate.

{}From Figs. 10 and 11, it is apparent that the periodic instanton
solutions interpolate smoothly between the perturbative region and the
sphaleron. In particular, the energy monotonically increases with
$\beta$ for the entire range of $\beta$, $[0,\beta_-]$ in accordance 
with (\ref{sec:dbde}).
 
%PUT THE NUMERICAL RESULTS HERE (FIGS10 - 12)
%PUT THE CONFIGS HERE PLUS COMP. WITH PERT. (FIGS13-14)
\vspace{.4cm} 
\epsfxsize=6cm
\epsfysize=5cm
\centerline{\epsfbox{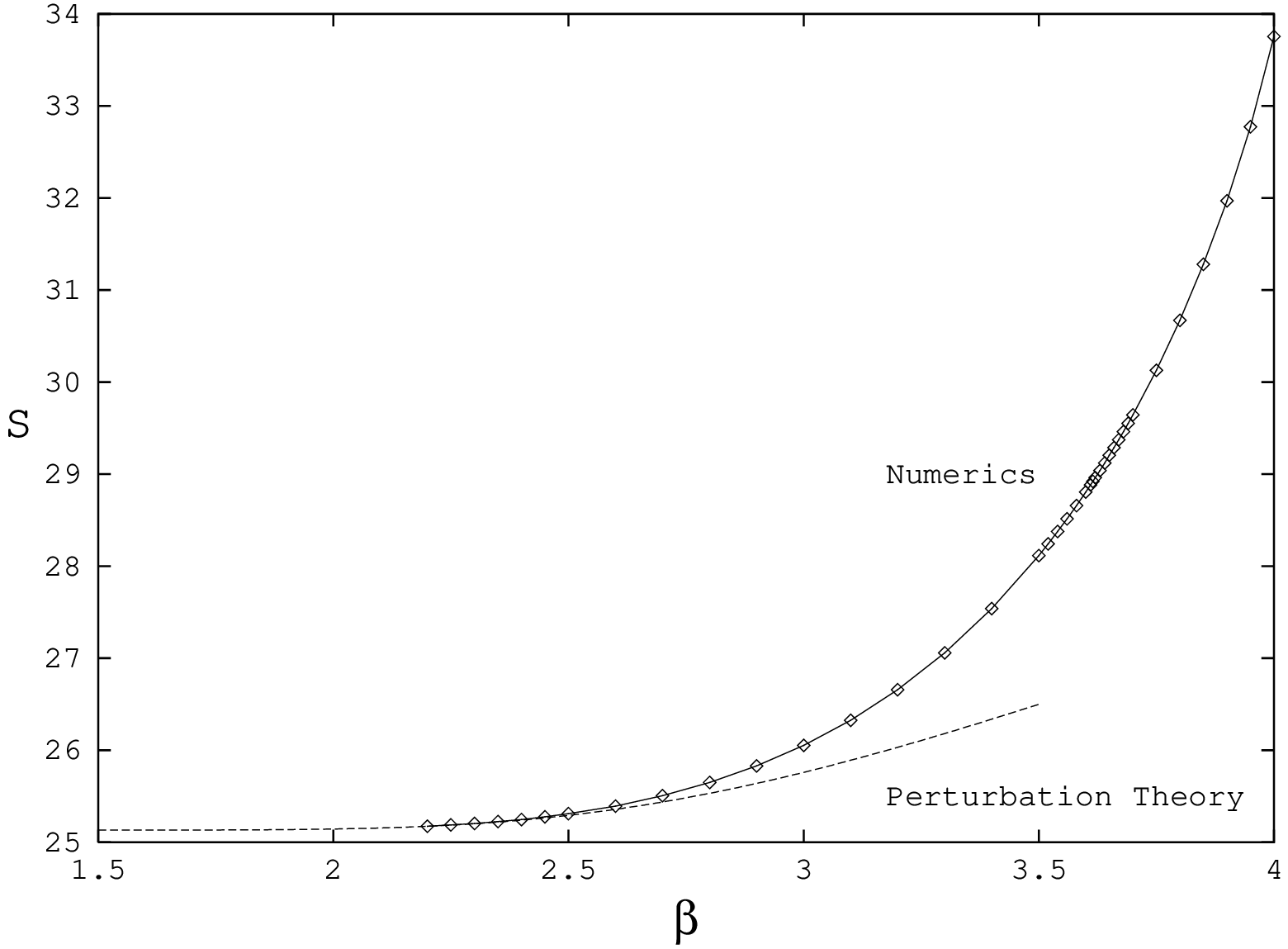}} 
\vspace{.35cm}
{FIG. 10. {\small{(a) The action of the periodic instanton
as a function of $\beta$. Agreement between the numerics and the
perturbation theory of the previous section is excellent in the
expected range.}}}\\

\vspace{.35cm} 
\epsfxsize=6cm
\epsfysize=5cm
\centerline{\epsfbox{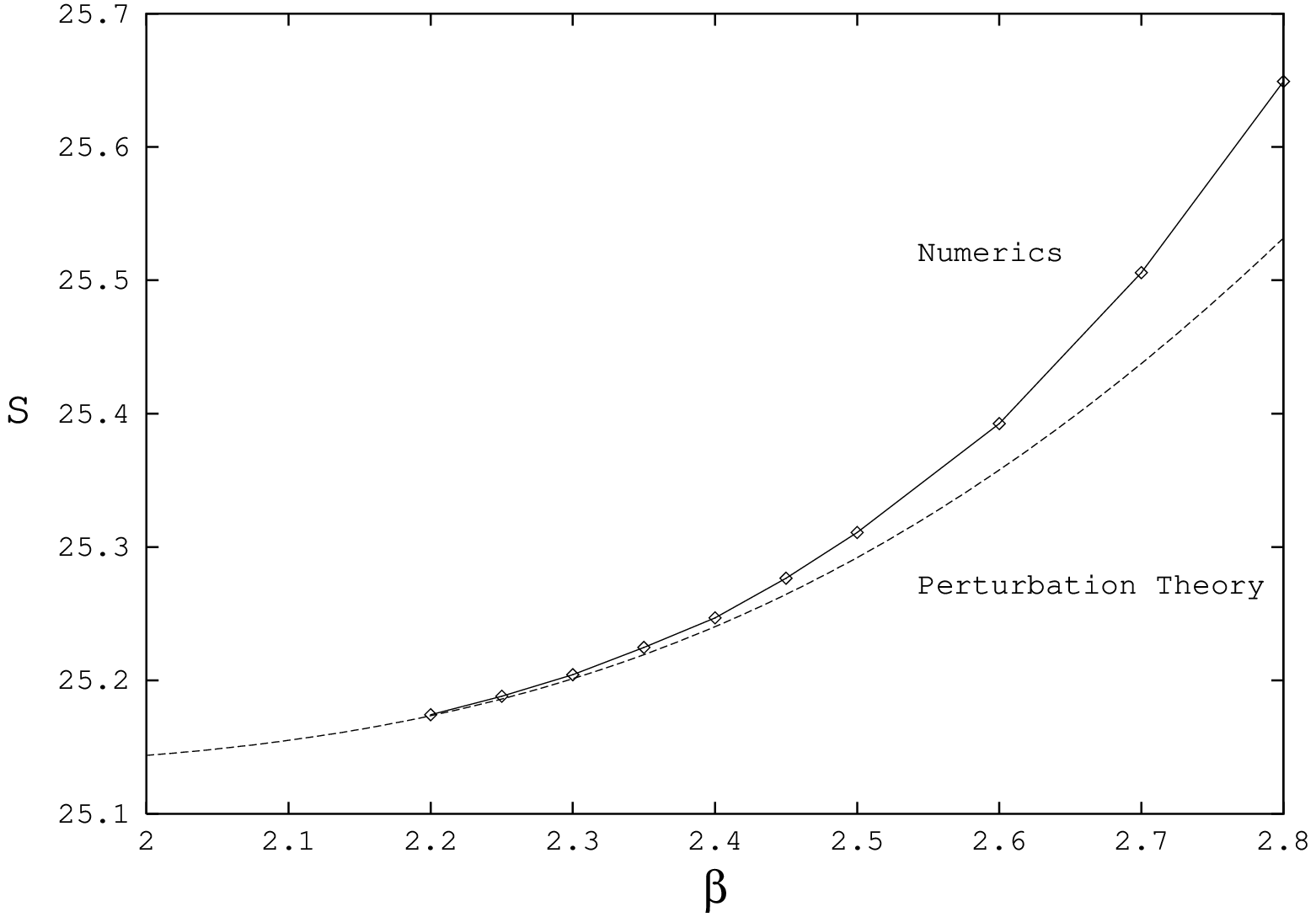}} 
\vspace{.35cm}  
{FIG. 10. {\small{(b) The same as Fig. 10(a) but with a more
restricted range for $\beta$ in order to better display the comparison
with the low energy perturbative expansion (dashed curve) of
Eqn. \ref{sec:Spinst}.}}}\\ 

\vspace{.35cm} 
\epsfxsize=6cm
\epsfysize=5cm
\centerline{\epsfbox{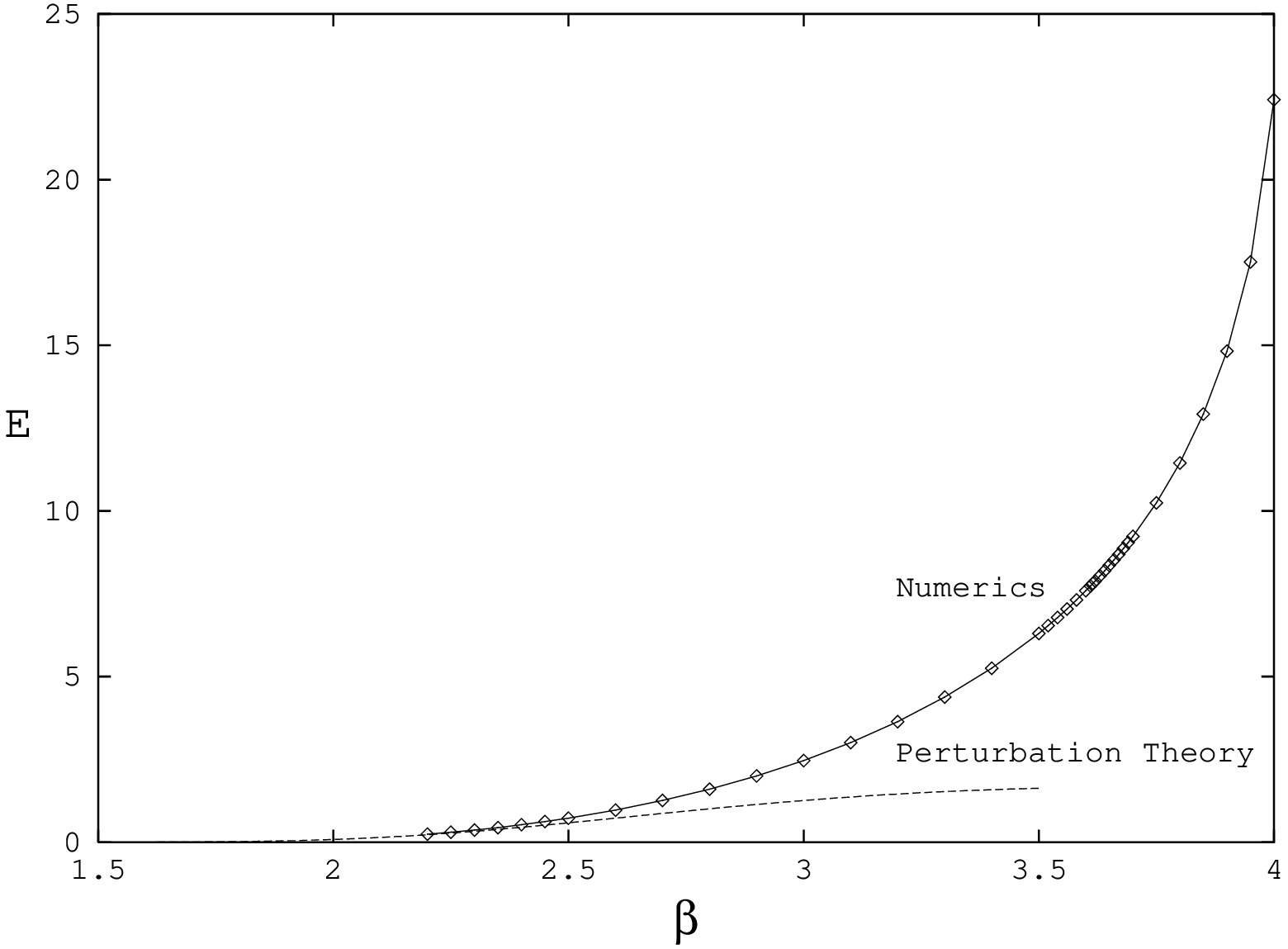}} 
\vspace{.35cm}
{FIG. 11. {\small{(a) The energy of the periodic instanton as
a function of $\beta$. The dashed curve is the perturbative result, Eqn. 
\ref{sec:lowE}.}}}\\

\vspace{.35cm} 
\epsfxsize=6cm
\epsfysize=5cm
\centerline{\epsfbox{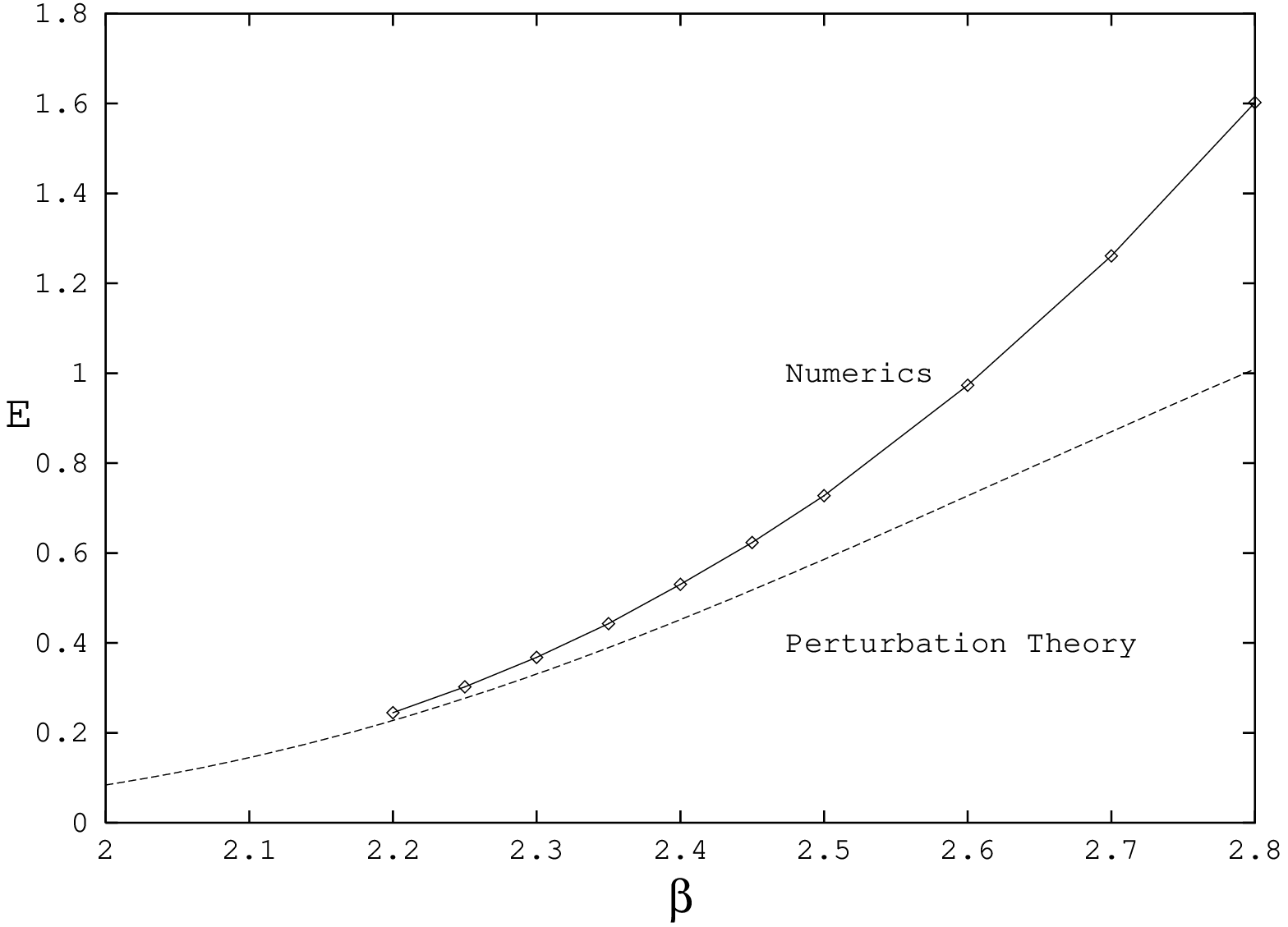}} 
\vspace{.35cm}
{FIG. 11. {\small{(b) The same as Fig. 11(a) but over a more
restricted range of $\beta$.}}}\\

\vspace{.35cm} 
\epsfxsize=6cm
\epsfysize=5cm
\centerline{\epsfbox{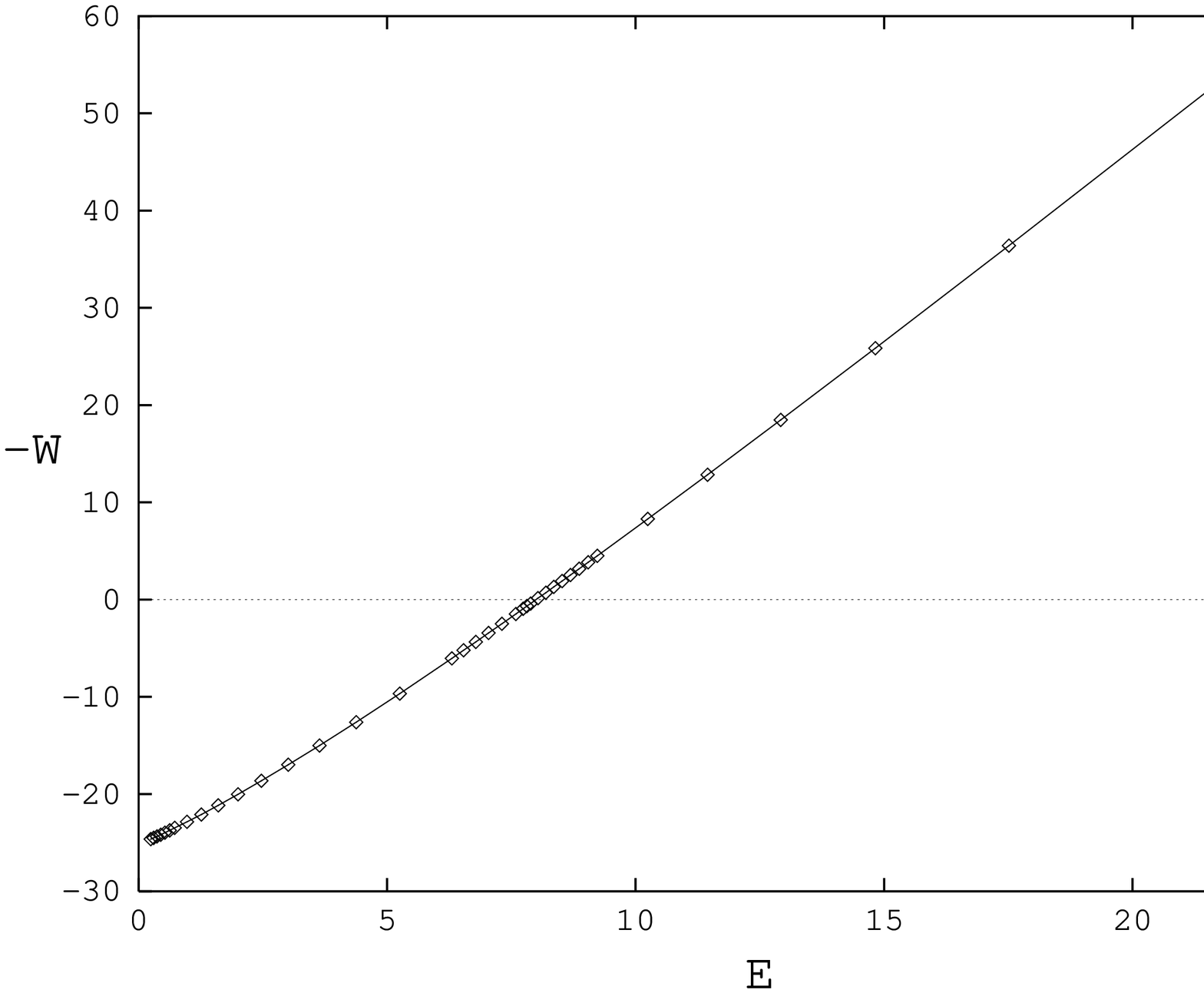}} 
\vspace{.35cm}
{FIG. 12. {\small{(a) -W as a function of the energy E. The 
change of sign in W occurs at $E=E_{sph}=8$.}}}\\  

\vspace{.35cm} 
\epsfxsize=6cm
\epsfysize=5cm
\centerline{\epsfbox{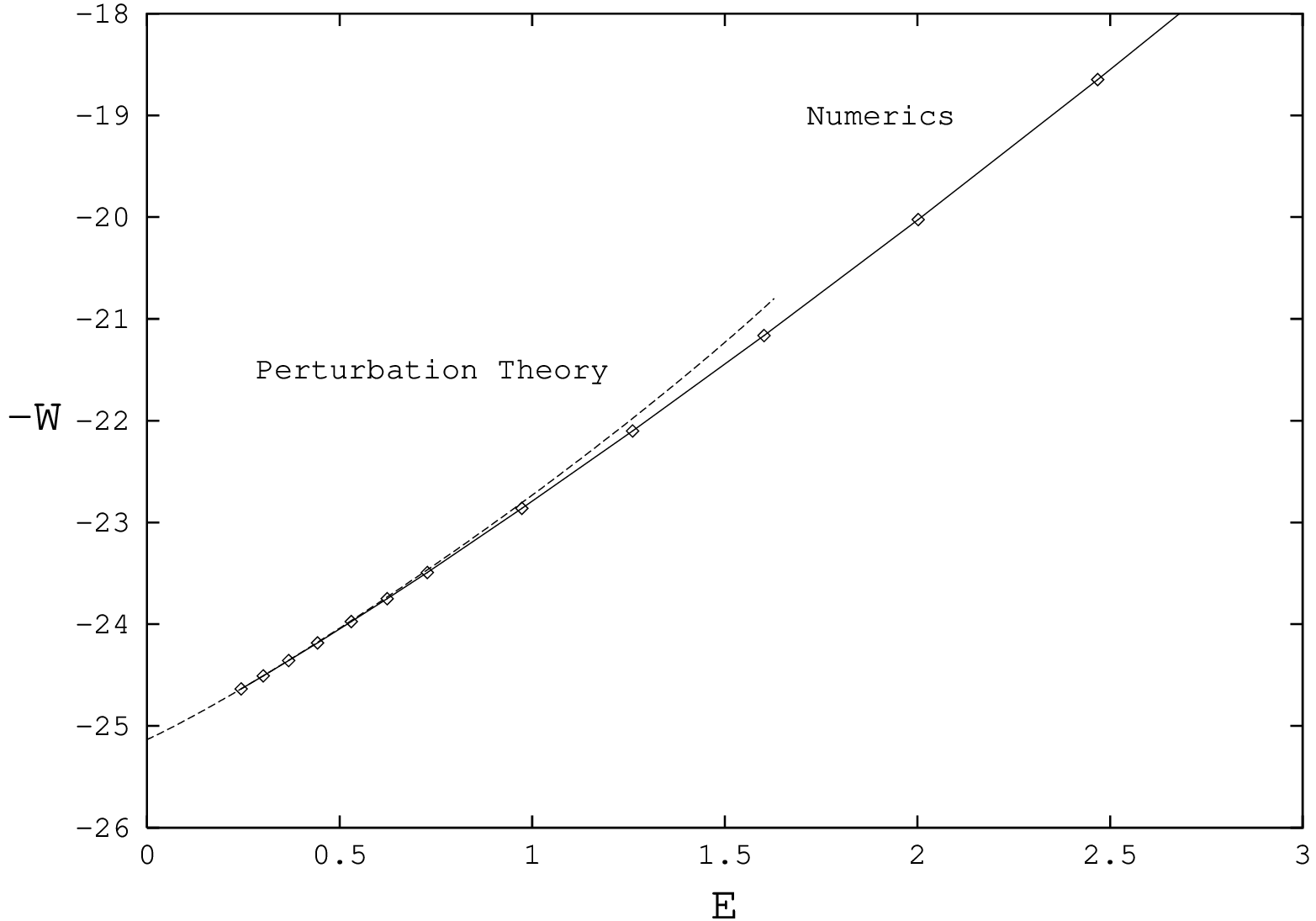}} 
\vspace{.35cm}  
{FIG. 12. {\small{(b) The same as Fig. 12(a) but over a more
restricted range of $\beta$ in order to display the comparison with
the perturbative result (dashed line).}}}\\ 

The numerically evaluated periodic instanton configurations are
displayed in Figs. 13 at $\beta=2.8$. These configurations go over to
the vacuum solution at the boundary $x=-L/2$ and are strongly non-vacuum 
near $\tau=\beta/4$, $x=0$. As $\beta$ is lowered, the periodic instanton 
size goes rapidly to zero (exponentially in the perturbative regime). At 
lower values of $\beta$, the configurations remain qualitatively
similar, but become ever more localized as functions of time and space.

The numerically computed periodic instantons can be compared to the
perturbative solution (\ref{sec:Bes}) with the instanton size $\rho$
given by (\ref{sec:lowErho}). Since direct comparison is unwieldy and
difficult to visualize, we chose to compare the numerical and
perturbative values for $|w|$ as a function of $\tau$ at $x=0$. This
comparison at $\beta=2.2$ is shown in Fig. 14. As is to be expected,
the overall agreement at this low value of $\beta$ is good. Evidently,
perturbation theory underestimates the interactions in the dilute gas
chain: the numerical solution has a longer tail and is less sharp than
the perturbative configuration near $\tau=\beta/4$.
\newpage
\vspace{-2cm}
\vspace{.4cm} 
\epsfxsize=9cm
\epsfysize=8cm
\centerline{\epsfbox{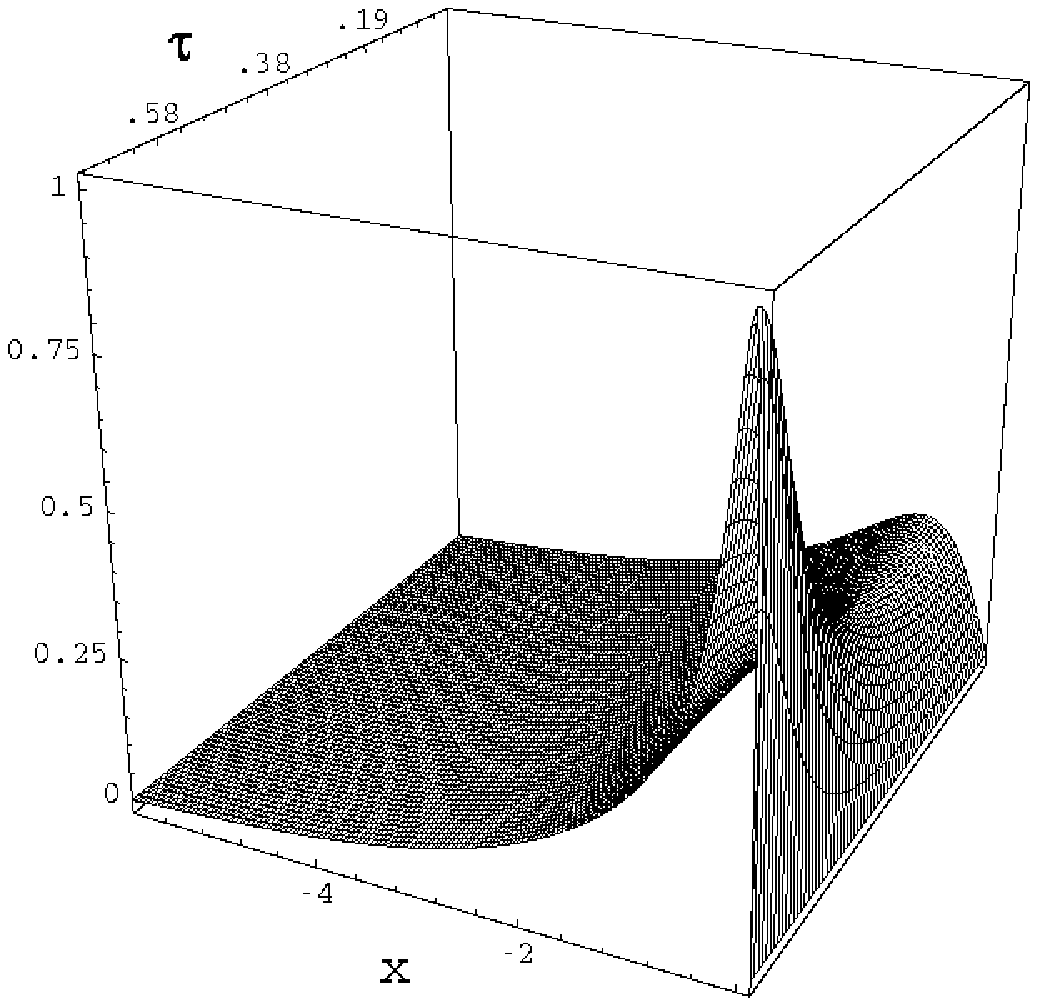}}
\vspace{-2cm}
\vspace{.5cm} 
\epsfxsize=9cm
\epsfysize=8cm
\centerline{\epsfbox{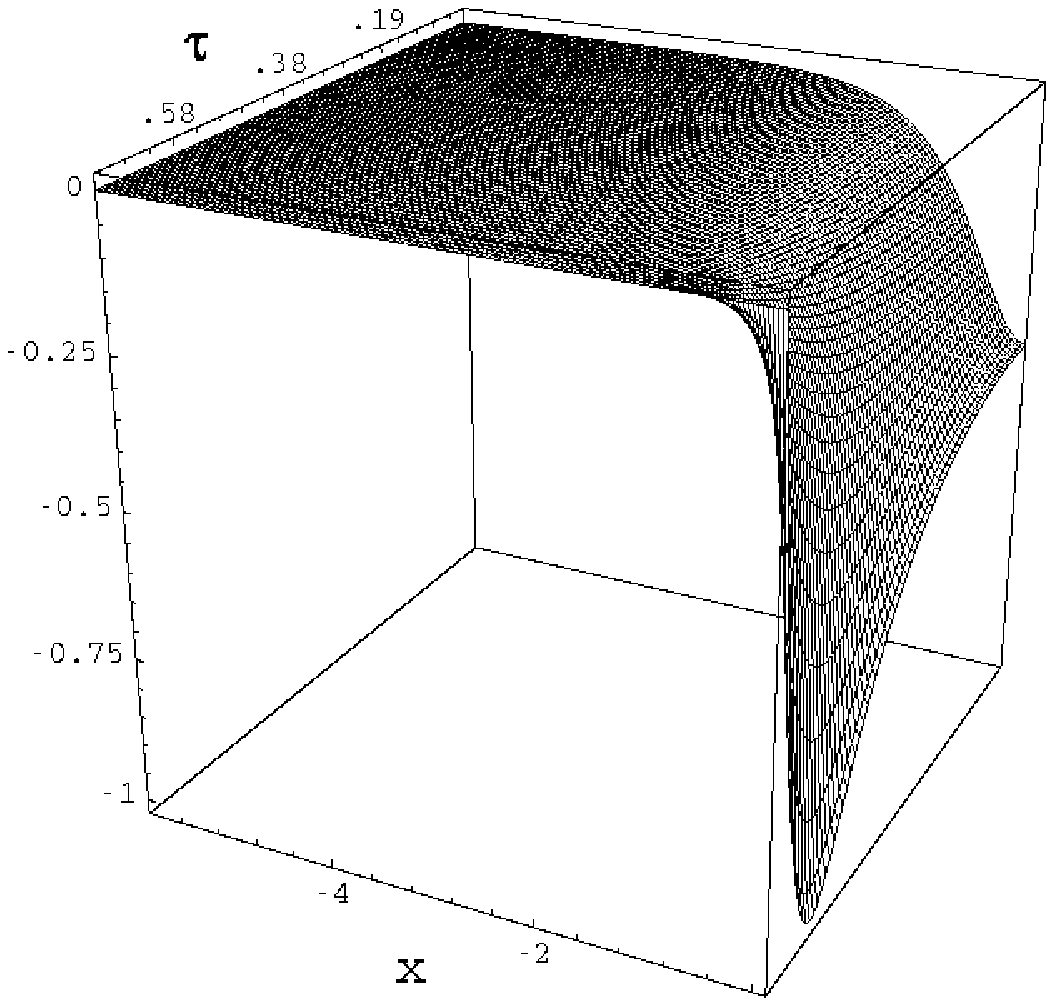}}
\vspace{-2cm}
\vspace{.5cm} 
\epsfxsize=9cm
\epsfysize=8cm
\centerline{\epsfbox{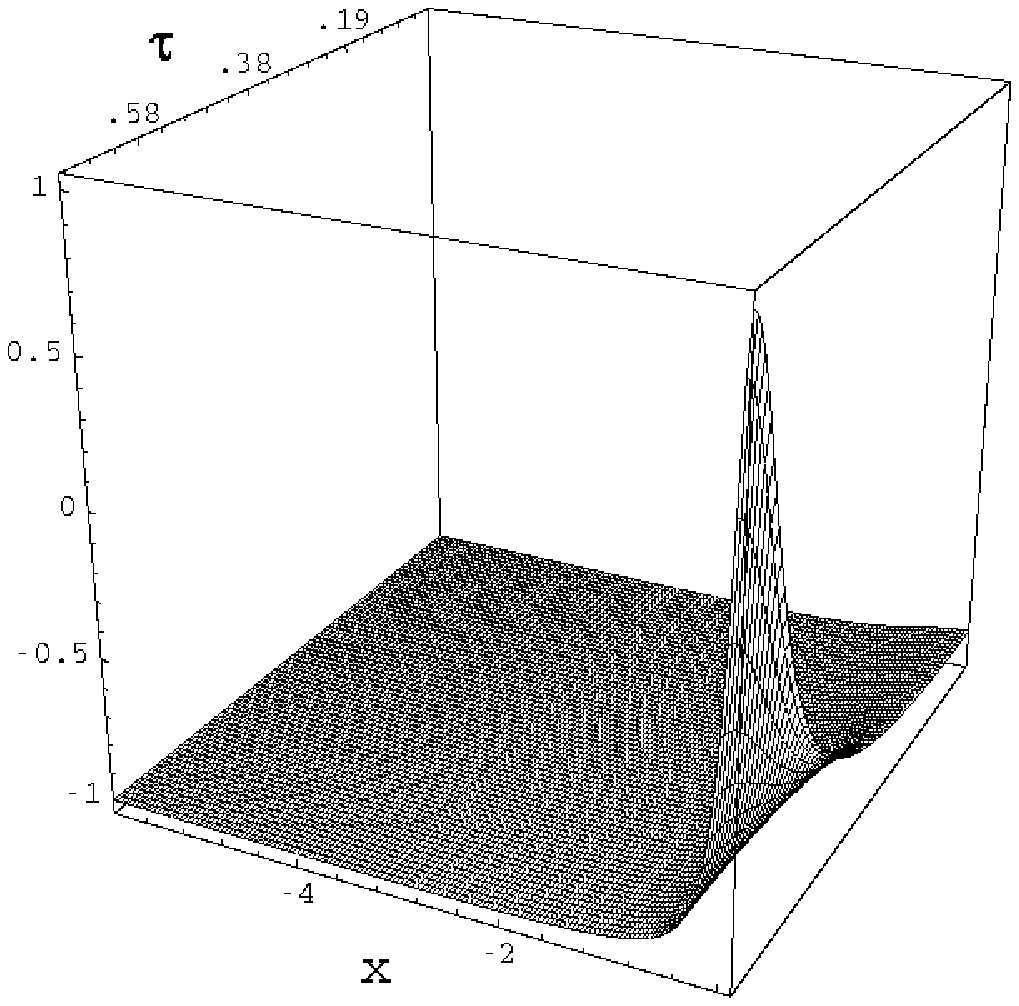}}
\vspace{-1cm}
{FIG. 13. {\small{A typical numerically obtained periodic 
instanton solution (here at $\beta = 2.8,~L=12$). Shown in the three
figures are the Cartesian components of the unit vector $n^1$, $n^2$
and $n^3$ as functions of $\tau$ and $x$ on a $73\times 153$ grid
($\tau\times x$). As one can see, at the turning point $\tau=0$ 
the field configuration is non-vacuum, and the components of the 
vector $n^a$ cover the shaded patch of the sphere in Fig. 7.}}}\\

\vspace{.5cm} 
\epsfxsize=6cm
\epsfysize=5cm
\centerline{\epsfbox{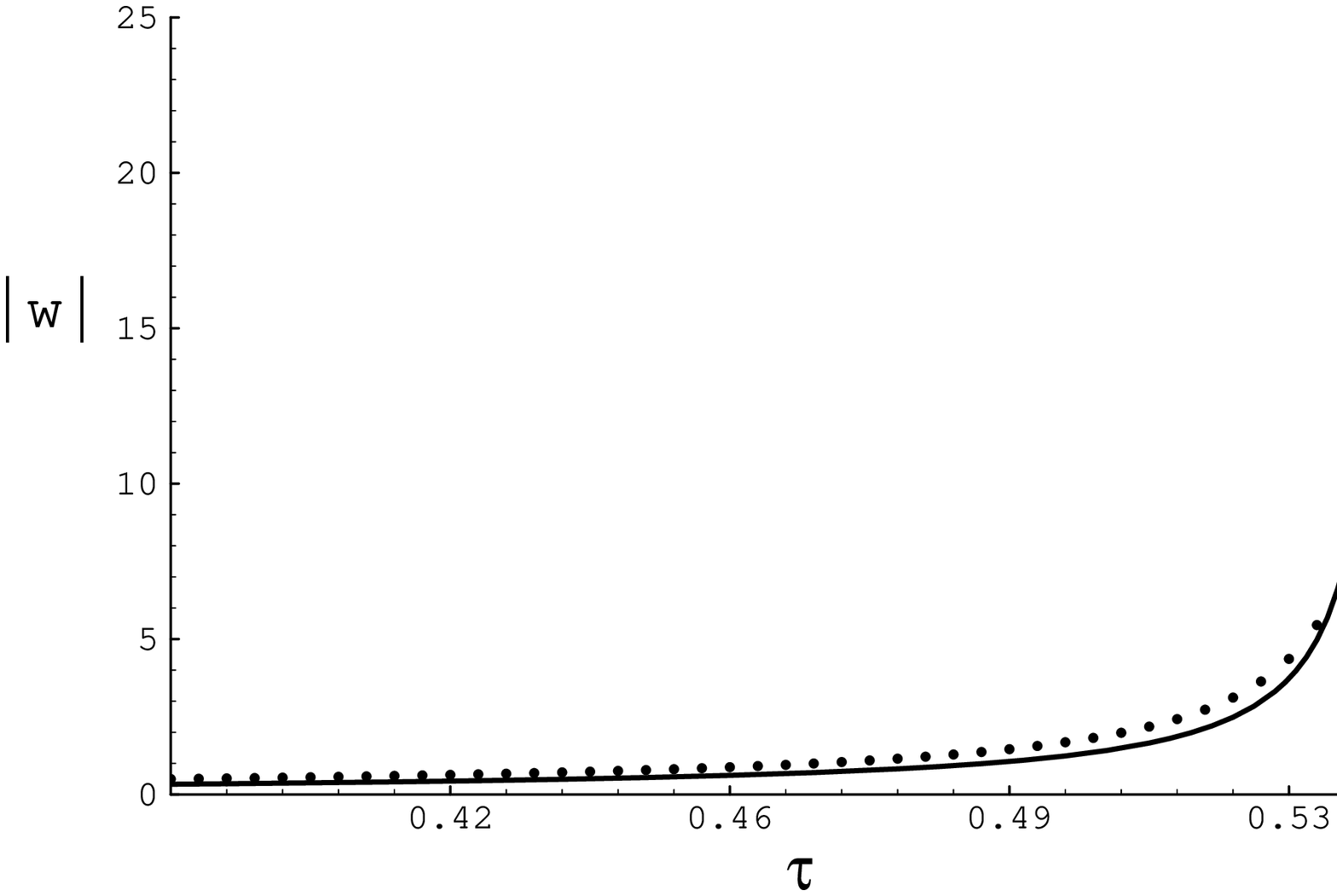}}
\vspace{.35cm}
{FIG. 14. {\small{Comparison of the perturbative (solid line)
and numerical results (dots) for $|{w}|$ as a function of $\tau$ at
$x=0$, $\beta=2.2$. Only a small piece of the lattice near
$\beta/4=.55$ is shown for clarity.}}}\\

\section{Summary and Discussion}
\label{sec:level6}

The main conclusion of our study of periodic instanton solutions in
the quantum pendulum and the $O(3)$ sigma model is that two
qualitatively and quantitatively different behaviors are possible for
these solutions, with correspondingly different physical consequences
for finite energy and temperature transitions.

In the first case (I)
of which the pendulum is the prototype, the period $\beta$ decreases
with increasing energy, $E_{sph}/T_- < 2S_0$, the periodic instanton
solutions contribute to the finite temperature winding number
transition rate, and the crossover from quantum tunneling to thermal
activation is smooth:
\begin{equation} 
{d E\over d \beta} < 0\,; \qquad {E_{sph}\over T_-} < 2S_0~. \qquad
{\rm case\ (I)} 
\label{sec:caseI}
\end{equation}
In addition to the periodic pendulum which we have discussed
here in some detail, the symmetric double well potential,
\begin{equation}
V(q) = {\omega^2 \over 8}(q^2 -1)^2
\end{equation}
belongs in this class, as may be easily checked from $E_{sph} = V(0) =
\omega^2/8$, $T_- = |{V''(0)}|^{1\over 2}/2\pi =
\omega/2\sqrt{2}\pi$, 
and the kink action $S_0 = 2\omega/3$. Hence, 
\begin{equation}
{E_{sph}\over T_-} = {\pi \sqrt{2} \over 4}\omega < 2 S_0 = {4\over
3}\omega\,, 
\end{equation}
and indeed the periodic instanton solutions with the expected behavior
are easily found analytically in this example as well. 

In two dimensions the most well-studied model possessing both instanton 
and sphaleron solutions is the abelian Higgs' model, {\em viz.}
\begin{equation}
S = \int d^2x \left\{ {1\over 4} F_{\mu\nu}^2 + \vert
D_{\mu}\Phi\vert^2 + \lambda\left(\vert\Phi\vert^2 - {1\over 2}
v^2\right)^2\right\}\,. 
\end{equation}
It is well-known that the instanton solution of this model is an
Abrikosov-Nielsen-Olsen vortex of the form \cite{ANO},
\begin{eqnarray}
A_{\mu} &=& {1\over g} \epsilon_{\mu\nu} {x^{\nu}\over r^2}
A(r)\nonumber\\ 
\Phi &=& v e^{i\phi} H(r)\,.
\end{eqnarray}
In the special case that $M^2_H = 2\lambda v^2$ is taken equal to
$M_W^2 = g^2 v^2$, the vortex may be found analytically, and has the
action, $S_0 = \pi v^2$.  For other values of the ratio $M_H/M_W$ the
vortex solution has to be found numerically \cite{JacReb}. The
sphaleron and its negative frequency mode are both known analytically
for arbitrary values of this ratio \cite{Boch}, {\em viz.}
\begin{eqnarray}
E_{sph} &=& {2\over 3}\sqrt{2\lambda} v^3\, \qquad {\rm and}\nonumber\\
\epsilon_-^2 &=& - {\lambda v^2 \over 4} \left( \sqrt{ 1 + {8g^2\over
\lambda}} + 1\right)\,. 
\end{eqnarray}
Hence at least in the case $g^2 = 2\lambda$ we find
\begin{equation} 
{E_{sph}\over T_-} = {8\pi \sqrt{2}\over 3 (\sqrt{17} +1)} v^2 < 2S_0
= 2\pi v^2 
\label{sec:testI}
\end{equation}
and the two dimensional abelian Higgs' model also falls into case (I).
The periodic instanton solutions have been found numerically in this
model for $M_W = M_H$ \cite{MJ}, and they do satisfy
(\ref{sec:caseI}). In accordance with the intuition gained by our
study of the sigma model, this behavior is the result of the absence
of any conformal invariance or scale parameter akin to $\rho$ in the
instanton solutions of the abelian Higgs' model, and we would expect
this model to behave in the same way for all finite values of the
ratio $M_H/M_W$. The numerical results for the vortex action in the
literature are consistent with (\ref{sec:testI}) for all values of
this ratio.

The $O(3)$ sigma model is the only example studied so far in any
detail which behaves differently, {\em i.e.}
\begin{equation} 
{d E\over d \beta} > 0\,; \qquad E_{sph}/T_- > 2S_0~. \qquad {\rm
case\ (II)} 
\label{sec:caseII}
\end{equation}
In this second case (II), the periodic instanton solutions contribute
to the finite {\em energy} (but {\em not} the finite temperature)
winding number transition rate, and the crossover from quantum
tunneling to thermal activation is sharp, taking place at $T_{cr} >
T_-$. As we have seen the key physical difference between cases (I)
and (II) appears to be conformal invariance and the scale parameter
$\rho$ of the instanton configurations.

These same features are shared by the $O(3)$ sigma model and four
dimensional gauge theories with spontaneous (and conformal) symmetry
breaking arising from the Higgs' sector. Indeed, it is interesting to
compare the numerical results for the sphaleron energy and negative
frequency eigenvalue \cite{Yaf},
\begin{equation}
\begin{array}{cccc}
{M_H \over M_W} & {E_{sph}\over M_W/\alpha_W} & {-\omega^2\over M_W^2}
& {E_{sph}\alpha_W\over T_-}\\ 
0.0 & 3.0405 & 1.318 & 16.64\\
0.1 & 3.1384 & 1.486 & 16.18\\
1.0 & 3.6417 & 2.460 & 14.59\\
2.0 & 3.9532 & 3.257 & 13.76\\
3.0 & 4.1633 & 3.967 & 13.13\\
4.0 & 4.3179 & 4.667 & 12.59\\
5.0 & 4.4375 & 5.405 & 11.99\\
7.0 & 4.6115 & 7.176 & 10.82\\
10.0& 4.7805 & 11.21 &  8.97
\end{array}
\nonumber
\end{equation}
to twice the zero energy instanton action $2 S_0$ of the pure
$SU(2)$-Higgs doublet model in four dimensions (which is the standard
electroweak model with $\theta_W =0$ and no fermions).  Since
$2S_0\alpha_W = 4\pi = 12.566\dots$ we observe from the last column of
the table above that $E_{sph}/T_- > 2 S_0$ for all $M_H / M_W < 4.0$
but that the inequality is reversed for all $M_W$ greater than a value
slightly larger than $4M_W$. Hence we conclude that the four
dimensional $SU(2)$-Higgs' doublet model falls under case (II) for
$M_H < 4M_W$.

Based on our study of instanton solutions in this paper we are led to
suspect that the four dimensional gauge theory is similar to the
$O(3)$ model, for not too large Higgs' self-coupling. Indeed, since
$SU(2)$ pure gauge theory is conformally invariant, there is an
instanton scale parameter $\rho$, which is driven to zero by the
addition of the Higgs' sector with its conformal breaking expectation
value $v$. Hence the instanton or anti-instanton solution no longer
exists in isolation for non-zero Higgs' mass $M_H = \sqrt
{2\lambda}v$.  However, the interaction between instantons and
anti-instantons is again attractive, so that at finite periodicity
$\beta$ there is an opposing interaction, and it is possible to
balance the two interactions at a definite value of $\rho(\beta)$,
just as in the $O(3)$ model considered in this paper. As in the sigma
model, in the spontaneously broken $SU(2)$ gauge theory the period of
the periodic instanton increases with energy at low energies, where
perturbation theory is applicable \cite{KRT}. Thus we conclude that
the four dimensional $SU(2)$-Higgs gauge theory falls into case (II)
for not too strong Higgs' self-coupling, and has a sharp crossover at
$T_{cr} > T_-$ from quantum tunneling to thermal activation in its
winding number transition rate.

Apparently the behavior changes at large $M_H/M_W$, as the model
becomes more ``Higgs-like'' and the instanton solutions resemble the
scale invariant $SU(2)$ instantons less and less. At Higgs'
self-coupling $\lambda \ge 8\pi \alpha_W$ (approximately) the
$SU(2)$-Higgs doublet model would appear to lie in case (I). At
higher values of $\lambda$ additional static sphaleron solutions
bifurcate from the simplest spherically symmetric one, and the
semiclassical approximation which requires $\lambda \ll 1$ eventually
ceases to be valid \cite{Yaf}. These results have been obtained at
zero $U(1)$ mixing angle $\theta_W$. Although sphaleron solutions have
been constructed at non-zero $\theta_W$ \cite{Kunz}, the negative
eigenmode and eigenvalue have not been given in the literature to our
knowledge, and therefore we are not yet able to draw any firm
quantitative conclusions in the case of non-zero $\theta_W$. However,
the behavior of the bosonic sector of the standard model with
$\theta_W$ near its physical value is not expected to be very
different from the behavior at zero $\theta_W$. In any case, we expect
that the physical electroweak theory behaves like the $O(3)$ model for
low to moderate $M_H/M_W$ in case (II) but the situation with the
additional Higgs coupling $\lambda$ in $3+1$ dimensions is more
complicated and deserving of detailed investigation.
      
To conclude, the classical solution space of even relatively simple
models such as the $O(3)$ model we have studied in this paper appears
to be very rich, and the more intricate details of the periodic
solutions to four dimensional spontaneously broken Yang-Mills theories
remain to be investigated. Both the analytic and numerical approaches
applied here to the $O(3)$ model will extend to the electroweak theory
in a straightforward way, and we believe that this program is worth
carrying through to completion. Besides the intrinsic interest of new
solutions of the Euler-Lagrange equations of a physical gauge theory,
this study, possibly extended into the complex domain appears to be
the only viable method at our disposal to investigate anomalous baryon
number violation at finite energy in the standard model.

\section{Acknowledgments}

The authors are grateful to T. Bhattacharya, M. Mattis,
M. E. Shaposhnikov and V. A. Rubakov for helpful
discussions. S. H. thanks Eugene Lo of Thinking Machines for advice on
the numerics. P. T. wishes to thank D. T. Son, A. Kuznetsov, and
S. Yu. Khlebnikov for useful discussions, and the Theoretical Division
of LANL for hospitality. S. H. and E. M. acknowledge support from the
United States Department of Energy. The work of P. T. is supported in
part by the Russian Foundation for Fundamental Research (grant
96-02-17804a) and INTAS (grant INTAS-94-2352). Numerical simulations
were performed on the CM-5 at the Advanced Computing Laboratory (ACL),
Los Alamos National Laboratory.

\end{document}